\documentclass[manuscript]{acmart}

\AtBeginDocument{%
  }

\setcopyright{acmlicensed}
\copyrightyear{2018}
\acmYear{2018}
\acmDOI{XXXXXXX.XXXXXXX}

\acmJournal{TOSEM}
\acmVolume{37}
\acmNumber{4}
\acmArticle{111}
\acmMonth{8}

\usepackage{tikz}
\usepackage{tabularx}
\usepackage{booktabs}
\usepackage{multirow}
\usepackage{makecell}
\usepackage{colortbl}
\usepackage{tcolorbox}
\tcbuselibrary{most, skins}
\usepackage{setspace}
\usepackage{pifont}
\usepackage{ragged2e}
\usepackage{xcolor}
\usepackage{url}
\usepackage{graphicx}

\newcolumntype{P}[1]{>{\raggedright\arraybackslash}p{#1}} 

\begin{document}
\title{PseudoBridge: Pseudo Code as the Bridge for Better Semantic and Logic Alignment in Code Retrieval}


\author{Yixuan Li}
\email{yxli24@m.fudan.edu.cn}
\orcid{0000-0002-9229-7555}
\affiliation{
  \institution{Shanghai Key Laboratory of Data
Science, School of Computer Science, Fudan University}
  \city{Shanghai}
  \country{China}
}

\author{Xinyi Liu}
\email{liuxiny24@m.fudan.edu.cn}
\orcid{0009-0007-1893-4417}
\affiliation{%
  \institution{Shanghai Key Laboratory of Data
Science, School of Computer Science, Fudan University}
  \city{Shanghai}
  \country{China}
}

\author{Weidong Yang}
\authornote{Corresponding Authors}
\email{wdyang@fudan.edu.cn}
\orcid{0000-0002-6473-9272}
\affiliation{%
  \institution{Shanghai Key Laboratory of Data
Science, School of Computer Science, Fudan University}
  \city{Shanghai}
  \country{China}
}

\author{Ben Fei}
\email{benfei@cuhk.edu.hk}
\orcid{0000-0002-3219-9996}
\affiliation{%
  \institution{Department of Information Engineering, The Chinese University of Hong Kong}
  \city{Hong Kong}
  \country{China}
}

\author{Shuhao Li}
\email{shli23@m.fudan.edu.cn}
\orcid{0009-0008-5175-7667}
\affiliation{%
  \institution{Shanghai Key Laboratory of Data
Science, School of Computer Science, Fudan University}
  \city{Shanghai}
  \country{China}
}

\author{Mingjie Zhou}
\email{mjzhou19@fudan.edu.cn}
\orcid{0000-0002-3289-0533}
\affiliation{%
  \institution{Shanghai Key Laboratory of Data
Science, School of Computer Science, Fudan University}
  \city{Shanghai}
  \country{China}
}

\author{Lipeng Ma}
\authornotemark[1]
\email{lpma21@m.fudan.edu.cn}
\orcid{0000-0001-5974-5988}
\affiliation{%
  \institution{Shanghai Key Laboratory of Data
Science, School of Computer Science, Fudan University}
  \city{Shanghai}
  \country{China}
}

\renewcommand{\shortauthors}{Li et al.}

\begin{abstract}
Code retrieval aims to precisely find relevant code snippets that match natural language queries within massive codebases, playing a vital role in software development. 
Recent advances leverage pre-trained language models (PLMs) to bridge the semantic gap between unstructured natural language (NL) and structured programming languages (PL), yielding significant improvements over traditional information retrieval and early deep learning approaches. 
However, existing PLM-based methods still encounter key challenges, including a fundamental semantic gap between human intent and machine execution logic, as well as limited robustness to diverse code styles.
To address these issues, we propose \textbf{PseudoBridge}, a novel code retrieval framework that introduces pseudo-code as an intermediate, semi-structured modality to better align NL semantics with PL logic. 
Specifically, PseudoBridge consists of two stages: First, we employ an advanced large language model (LLM) to synthesize pseudo-code, enabling explicit alignment between NL queries and pseudo-code. Second, we introduce a logic-invariant code style augmentation strategy and employ the LLM to generate stylistically diverse yet logically equivalent code implementations with pseudo-code, then align the code snippets of different styles with pseudo-code, enhancing model robustness to code style variation. 
We build PseudoBridge across 10 different PLMs and evaluate it on 6 mainstream programming languages. 
Extensive experiments demonstrate that PseudoBridge consistently outperforms baselines, achieving significant improvements in generalization, particularly in zero-shot scenarios like Solidity and XLCoST. 
Extended evaluations using open-source LLMs and advanced embeddings confirm that these gains stem from PseudoBridge's intrinsic design, independent of specific closed-source models.
PseudoBridge achieves performance comparable to state-of-the-art embedding methods, highlighting the effectiveness of explicit logical and semantic alignment via pseudo-code as a robust solution for code retrieval.

\end{abstract}



\maketitle

\section{Introduction}

Code retrieval refers to the task of searching relevant code snippets from a large codebase in response to a natural language query~\cite{xia2017developers,sachdev2018retrieval,xie2023survey}. 
Natural language queries are typically unstructured textual descriptions, whereas code snippets are expressed in structured programming languages.
Consequently, the core challenge in code retrieval is to effectively align natural language (NL) with programming language (PL).
The emergence of large language models (LLMs) positions code retrieval as a key component for paradigms such as Retrieval-Augmented Generation (RAG)~\cite{yang2025empirical} and In-Context Learning (ICL)~\cite{li2023large}.
In these frameworks, code retrieval functions as the information bridge, with retrieved results serving either as repository-level code context to augment domain knowledge in RAG paradigms or as few-shot exemplars to demonstrate reasoning patterns in ICL scenarios~\cite{zhang2023repocoder,chen2024code}.
However, the effectiveness of these downstream tasks heavily relies on the precision and relevance of the retrieved results. 
Retrieving semantically mismatched or noisy code snippets can mislead the reasoning process of LLMs, leading to hallucinations or logical errors in generation. 
Therefore, enhancing the robustness of code retrieval in complex retrieval scenarios is critical. 
It not only significantly improves developer efficiency and experience but also serves as a cornerstone for downstream LLM performance, constituting a prerequisite for the effective deployment of LLMs in software engineering.

Early approaches to code retrieval are primarily based on information retrieval (IR) techniques~\cite{robertson2009probabilistic,bajracharya2014sourcerer,lv2015codehow}, which treat code as plain text and rely on keyword matching. 
However, such token-based methods depend on lexical co-occurrence statistics and struggle to effectively capture the semantic correspondence between natural language (NL) queries and code snippets~\cite{gu2018deepcs,xie2023survey}.
The rapid development of deep learning (DL) has driven significant breakthroughs in code retrieval, thanks to its strong feature extraction and representation learning capabilities~\cite{xie2023survey,liu2021codematcher,cheng2022csrs}. 
DL-based methods can be generally divided into two categories. 
The first category employs conventional neural architectures (e.g., RNN)~\cite{cheng2022csrs,liu2021codematcher}, where static word embedding models are used to achieve vectors for both NL and code snippets, then leverage neural networks to align their semantics with supervised training.
The second category leverages pre-trained language models (PLMs)~\cite{feng2020codebert,guo2020graphcodebert,guo2022unixcoder,li2022coderetriever,shi2023cocosoda}.
These approaches utilize transformer-based architectures pre-trained on large-scale code and natural language corpora to learn universal semantic representations across both modalities.
Compared to earlier approaches, PLM-based methods achieve state-of-the-art performance by harnessing the powerful semantic representation capabilities of PLMs. 
Moreover, fine-tuning these models on task-specific datasets with cross-modal alignment further enhances their ability for specific code retrieval scenarios.
As a result, PLM-based methods have become the dominant approach in modern code retrieval, where their effectiveness lies in their ability to align semantic consistency between NL and PL.

\begin{figure}[htbp]
    \centering
    \includegraphics[width=0.8\linewidth]{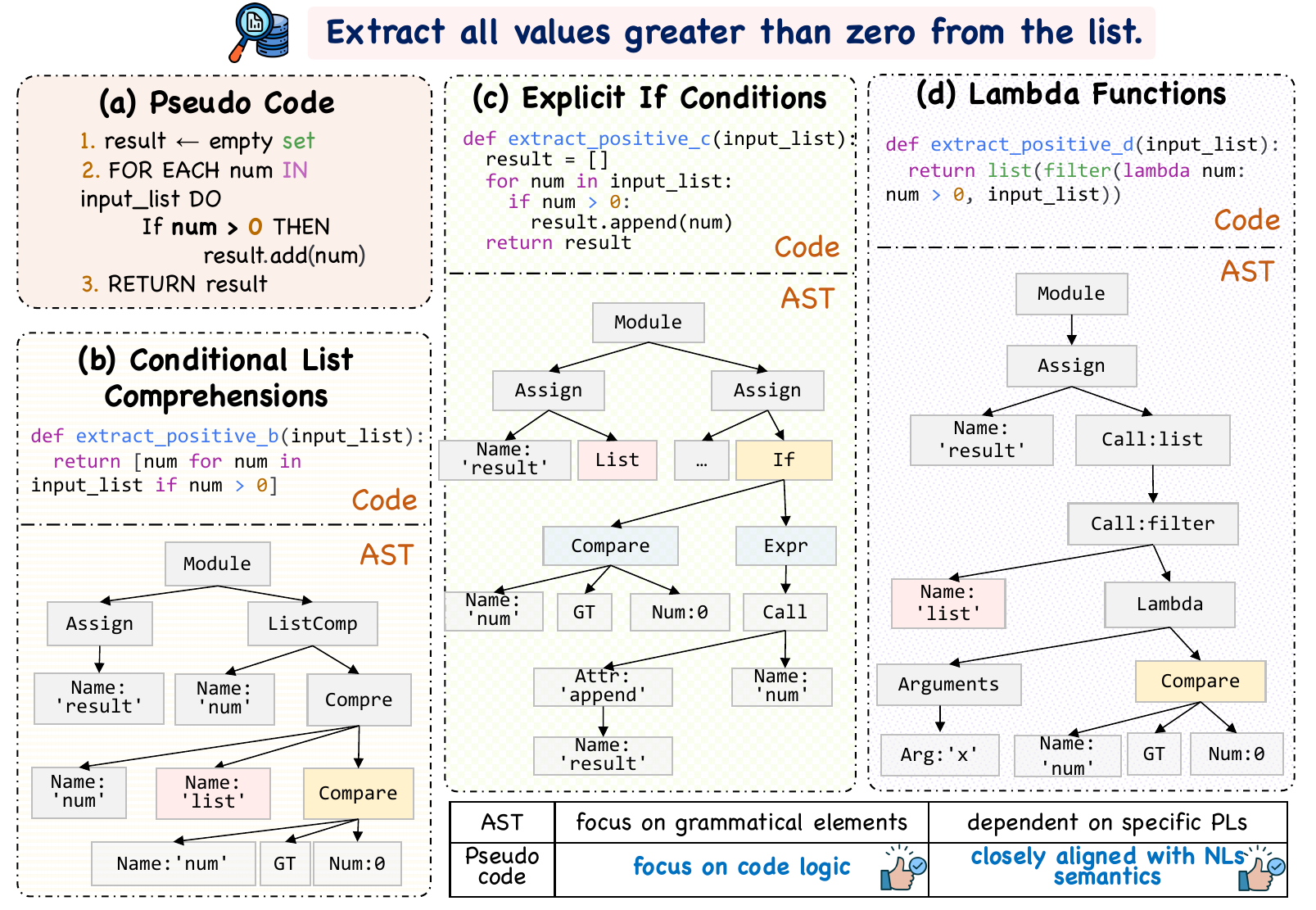}
    \caption{Comparison between Pseudo-Code and AST, where AST focuses on the structure of the code, not only verbose but also hard to express the logic.}
    \vspace{-0.15cm}
    \label{fig:compare_pc_ast}
\end{figure}

Although these PLMs methods for code retrieval have shown promising performance, they still face two fundamental challenges.
First, there remains a significant semantic gap between human intent and machine execution logic, which stems from the inherent differences between natural language and programming languages.
NL tends to be ambiguous, context-dependent, and rich in semantics, while PL is designed to be precise, unambiguous, and strictly structured to instruct computers.
For example, human concepts such as ``efficient'' do not have direct equivalents in PL, as they require an understanding of computational trade-offs like time or space complexity, and API optimizations. 
While PLMs have advanced cross-modal alignment~\cite{feng2020codebert}, they focus on surface-level semantic representations, making it difficult to bridge the inherent gap between semantics and logic.
Many works~\cite{guo2022unixcoder,premtoon2020semantic} attempt to address this issue by incorporating code structure information as additional supervision signals, particularly through Abstract Syntax Trees (ASTs), as illustrated in Figure \ref{fig:compare_pc_ast}.  
However, ASTs mainly capture the syntactic structure of code and introduce noise through their sheer volume and low-level granularity, lacking explicit modeling of the underlying program logic.
Consequently, this gap remains unaddressed, and models struggle to fully align the intent behind user queries with the logical information encoded in programs.

Second, current PLMs often overlook the impact of code style diversity. 
In practice, the same functionality can be implemented in multiple code style implementations, as shown in Figure~\ref{fig:compare_pc_ast}. 
For example, for the query requirement of ``extracting all numbers greater than zero from the list,'' there may exist several functionally equivalent code snippets: (b) list comprehensions that embed filtering logic, (c) explicit \texttt{if} conditions within loops, and (d) lambda functions combined with higher-order functions.
This phenomenon means that a single natural language query can correspond to various code snippets that share the same underlying logic but differ significantly in their syntactic structure and implementation style. However, current PLMs struggle to align these logically equivalent yet stylistically diverse code snippets, leading to significant performance degradation and inconsistency in retrieval results.
Moreover, as illustrated in Figure~\ref{fig:compare_pc_ast}, variations of different styles manifest as substantial differences in their corresponding ASTs, causing models to focus excessively on superficial implementation details rather than core logic. 
This tendency not only increases the model's sensitivity to code styles but also hinders its ability to capture deep code semantics. As a result, the models may misinterpret the semantic consistency between natural language queries and their possible code implementations, ultimately hindering the effectiveness of code retrieval models.

To address these challenges, we propose \textbf{PseudoBridge}, a novel code retrieval framework that introduces pseudo-code as an intermediate bridge to better align NL semantics with PL logic. Pseudo-code, situated between informal natural language and formal programming language, serves as a semi-structured modality that effectively models both semantics and logic.
Specifically, our framework consists of two stages: in the first stage, we employ advanced LLM (e.g., GPT-4o) to synthesize pseudo-code for each <NL, PL> pair, and leverage pseudo-code as an intermediate representation to narrow the gap between natural language query and pseudo-code. 
In the second stage, we introduce a logic-invariant code style augmentation strategy, where we leverage LLM to synthesize multiple style-different but logically equivalent code implementations from the constructed pseudo-code. This augmentation further aligns the pseudo-code representations with various code styles.
This strategy not only bridges the inherent gap between different language structures but also improves robustness to code style variation and generalization across unseen domains.

To evaluate the effectiveness of PseudoBridge, we fine-tune PseudoBridge with 10 different PLMs as the backbone
and conduct comprehensive experiments across 6 mainstream programming languages. 
Results demonstrate that PseudoBridge greatly improves retrieval performance, confirming the efficacy of explicitly constructing logical alignment with pseudo-code. 
Notably, under zero-shot settings, PseudoBridge exhibits strong performance on domain-specific datasets (e.g., Solidity~\cite{chai2022cross} and XLCoST~\cite{zhu2022xlcost}), highlighting its exceptional generalization capability and practical potential.

In summary, the major contributions of this paper are as follows:
\begin{enumerate}
    \item We propose a novel code retrieval framework, called PseudoBridge. To the best of our knowledge, PseudoBridge is the first work to explicitly leverage pseudo-code as an intermediate bridge specifically for the code retrieval task. PseudoBridge employs pseudo-code in a structured yet readable form to bridge the semantic gap effectively between ambiguous NL queries and precise PL implementations, significantly enhancing generalization across diverse retrieval scenarios.
    \item We develop a two-stage training framework that progressively aligns NL with PL via pseudo-code, incorporating both semantic consistency and logic invariance. In the first stage, we synthesize pseudo-code with an advanced LLM and align the NL with the pseudo-code representation. In the second stage, we introduce a code style augmentation strategy and align the pseudo-code with diverse code implementations. This approach directs the focus of the model towards the underlying semantics rather than surface syntax, strengthening the robustness to stylistic variation and improving cross-domain generalization.
    \item We conduct extensive experiments by fine-tuning PseudoBridge on 10 different PLMs and evaluate its effectiveness across 6 PLs. Experimental results show that PseudoBridge significantly improves code retrieval across all tested models and languages, demonstrating strong zero-shot generalization on specialized datasets like Solidity and XLCoST without training. These results confirm the efficacy of pseudo-code as an intermediate representation and underscore the framework's practical potential for real-world multilingual retrieval. 
\end{enumerate}

\section{Related Works}

\subsection{Code Retrieval Techniques}

Code retrieval is a fundamental technology in intelligent software engineering, as effective retrieval systems greatly improve development productivity. 
Existing approaches can be broadly classified into three categories: information retrieval (IR) methods, deep learning (DL) models, and pre-trained language models (PLMs).
Traditional IR-based approaches make code plain text, relying on lexical matching and keyword indexing techniques~\cite{hill2011improving, mcmillan2011portfolio,lv2015codehow}. 
While these techniques are efficient, they are inherently limited by their reliance on lexical similarity, which prevents them from capturing the deeper semantic relationships between NL and PL. 
This lexical dependency restricts their ability to bridge the semantic gap and model complex code semantics.
With advances in deep learning, researchers have begun to exploit large-scale datasets to model the relationships between NL and PL. 
DL-based approaches employ neural network architectures to map NL queries and code snippets into a shared vector space, thereby facilitating semantic alignment. 
For example, DeepCS~\cite{gu2018deepcs} utilizes recurrent neural networks to embed both queries and code within a common representation space, while CodeMatcher~\cite{liu2021codematcher} enhances retrieval accuracy through semantics-aware query expansion mechanisms.

More recently, pre-trained language models (PLMs) have become the mainstream approach to code retrieval~\cite{di2023code}. Leveraging Transformer architectures, these models acquire comprehensive programming knowledge from large-scale code repositories~\cite{zhang2017expanding, feng2020codebert, cao2021automated}. 
PLMs are pre-trained on corpora containing both code and NL queries to learn cross-modal semantic representations~\cite{guo2022unixcoder, arakelyan2022ns3, cambronero2019deep}, and are subsequently fine-tuned on task-specific datasets to further align NL and PL semantics, thus boosting retrieval performance~\cite{hu2023revisiting, wang2021codet5, wang2023codet5+}.
Furthermore, SPENCER~\cite{gu2025spencer} proposes a self-adaptive model distillation framework that synergizes bi-encoders and cross-encoders to optimize both retrieval accuracy and inference efficiency.
Despite these advancements, most existing methods still emphasize lexical-level semantic similarity, often overlooking the logical consistency of code. 
As a result, models may fail to fully capture the intent behind natural language queries and struggle to establish deep logical alignment between queries and code snippets. 
This limitation undermines their effectiveness in accurately retrieving target code in real-world scenarios.
To address this challenge, our work introduces \textbf{pseudo-code} as an intermediate representation, serving as a bridge to enhance both semantic and logical alignment between NL and PL.

\subsection{Data Synthesis for Code Intelligence}
The synthesis of high-quality data has played a crucial role in advancing code intelligence tasks, particularly in improving the performance of LLMs for code generation and understanding. 
Early research explored the use of simulated coding environments to create large-scale datasets by generating and validating programming problems and solutions~\cite{haluptzok2022language,shypulalearning}.
For instance, techniques such as Self-play~\cite{haluptzok2022language} and Chain-of-Thought~\cite{shypulalearning} enable generating diverse and correct code.
The development of instruction-following code LLMs has also greatly benefited from synthetic data. 
WizardCoder~\cite{luo2024wizardcoder} uses the Evol-Instruct technique to iteratively increase instruction complexity, creating training sets with complex multi-constraint problems. Similarly, Magicoder~\cite{wei2024magicoder} introduces OSS-Instruct, which uses open-source code snippets as seeds to formulate realistic programming problems. This ensures that the synthetic data retains the diversity of real-world software.

In the specific domain of code retrieval, data synthesis mitigates the scarcity of aligned query and code pairs. InverseCoder~\cite{wu2025inversecoder} leverages robust code to natural language generation to synthesize high-quality queries from raw code, effectively expanding training datasets through self-improvement. Furthermore, approaches like SyNeg~\cite{li2024syneg} utilize LLMs to create synthetic hard negatives that are lexically similar yet semantically distinct from positive samples. Training on these synthetic samples compels models to learn finer semantic discrimination and significantly improves performance on dense retrieval benchmarks~\cite{wang2022gpl}.
Despite these improvements, existing synthesis techniques primarily focus on surface-level textual augmentation or local semantic refinement, leaving the fundamental gap between abstract logic and concrete syntax largely unaddressed. While structured reasoning frameworks like SCoT (Structured Chain-of-Thought) \cite{li2025scot} have explored programming structures (such as sequential, branch, and loop) to guide LLMs, their primary objective is to enhance the logical consistency of the generative process during inference.

In contrast, PseudoBridge introduces a paradigm shift by utilizing pseudo-code not as a reasoning prompt for generation, but as a logic-invariant intermediate representation specifically engineered for cross-modal retrieval alignment. While SCoT emphasizes control-flow integrity to ensure execution accuracy, PseudoBridge abstracts away syntactic noise to provide a ``semantic anchor'' that bridges natural language queries and source code. By decoupling algorithmic logic from language-specific implementation details, our approach fundamentally enhances the robustness and generalization of retrieval models across diverse programming languages.

\begin{figure}[]
    \centering
    \includegraphics[width=0.6\linewidth]{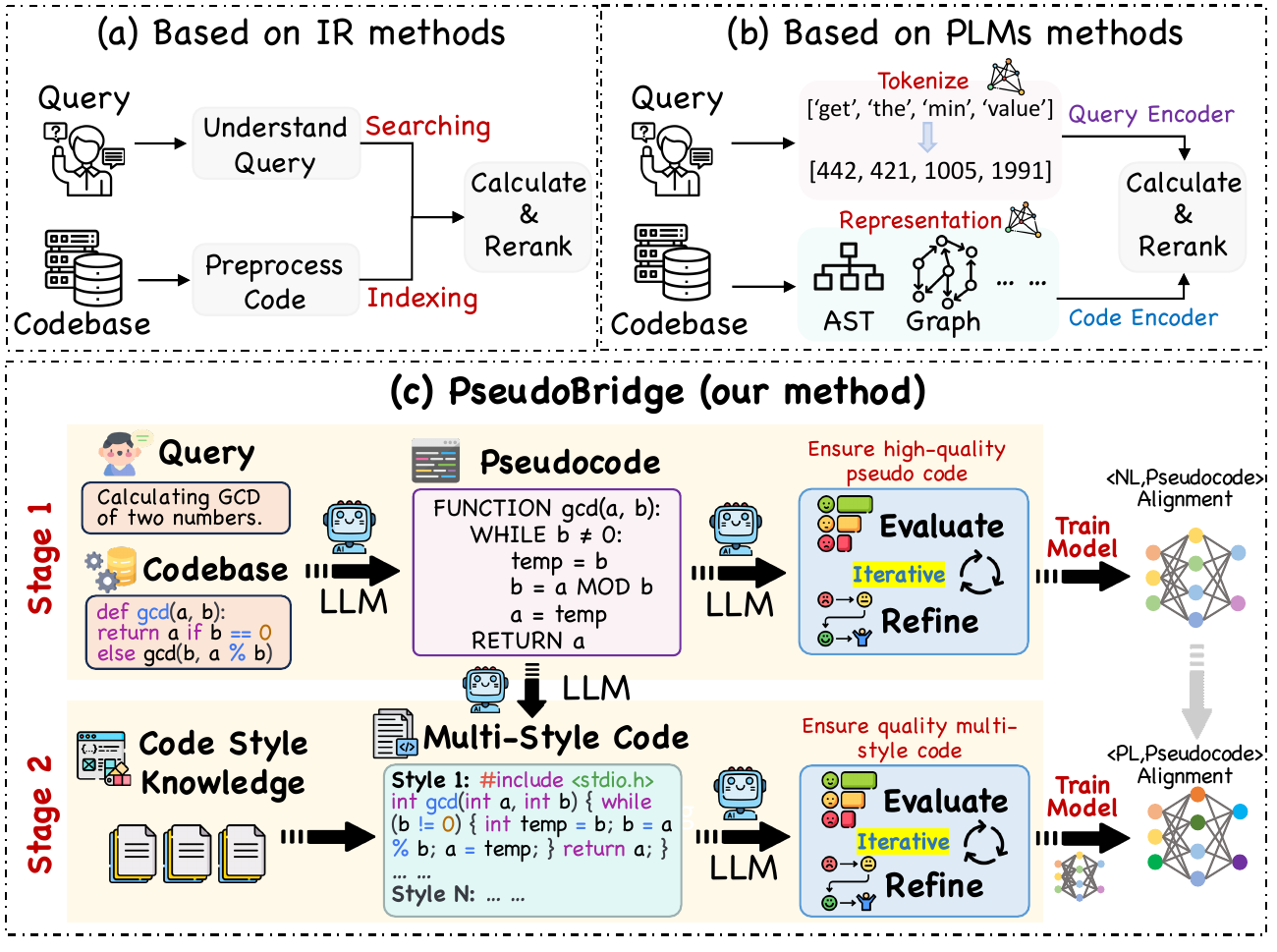}
    \caption{Comparison of Existing Code Retrieval Frameworks. IR-based and PLM-based methods achieve superficial alignment between code and queries. Our proposed framework utilizes pseudo-code and logic-invariant code style enhancement to better align NL semantics with PL logic.}
    \vspace{-0.2cm}
    \label{fig:methods_compare}
\end{figure}

\section{Methods}

\subsection{Overview}
The conceptual framework of PseudoBridge is depicted in Figure~\ref{fig:methods_compare} (c). 
In contrast to retrieval methods that rely on IR direct keyword matching (Figure~\ref{fig:methods_compare} (a)) or align NL and PL directly through PLMs (Figure~\ref{fig:methods_compare} (b)), PseudoBridge introduces a novel two-stage approach to more effectively align NL semantics with PL logic.
Specifically, PseudoBridge first employs pseudo-code as an intermediate representation to effectively narrow the semantic gap between human intent, expressed through natural language queries, and machine executable logic, represented by code snippets. 
This design facilitates the alignment and learning of core logic and semantics between the two modalities. 
Subsequently, PseudoBridge constructs diverse code variants to simulate realistic scenarios in which the same query corresponds to different code implementations. 
This strategy enhances the model's generalization capability and ultimately improves retrieval performance.
The framework operates in two distinct stages:

\textbullet~\textbf{Stage 1}: In this stage, PseudoBridge utilizes advanced LLMs to synthesize pseudo-code from each aligned <NL, PL> pair. The generated pseudo-code captures the core algorithmic logic and control flow described in the code, while maintaining a human-readable abstraction consistent with the original NL query. This pseudo-code acts as a semantic intermediary, allowing the model to better disentangle high-level intent from low-level implementation details.

\textbullet~\textbf{Stage 2}: Based on the synthesized pseudo-code, PseudoBridge employs the LLM to generate multiple code implementations that are functionally equivalent but syntactically diverse.
By aligning the pseudo-code representation with both the original and the augmented variants, the model learns to focus on invariant logical structures rather than incidental syntactic features. This strategy enhances the model’s robustness to code style variability and improves retrieval performance across heterogeneous codebases.

The following sections provide a detailed explanation of each component in PseudoBridge: pseudo-code generation (Section~\ref{sec3.2}), multi-style code generation (Section~\ref{sec3.3}), and model training (Section~\ref{sec3.4}).

\begin{figure*}[ht]
    \centering
    \includegraphics[width=0.95\linewidth]{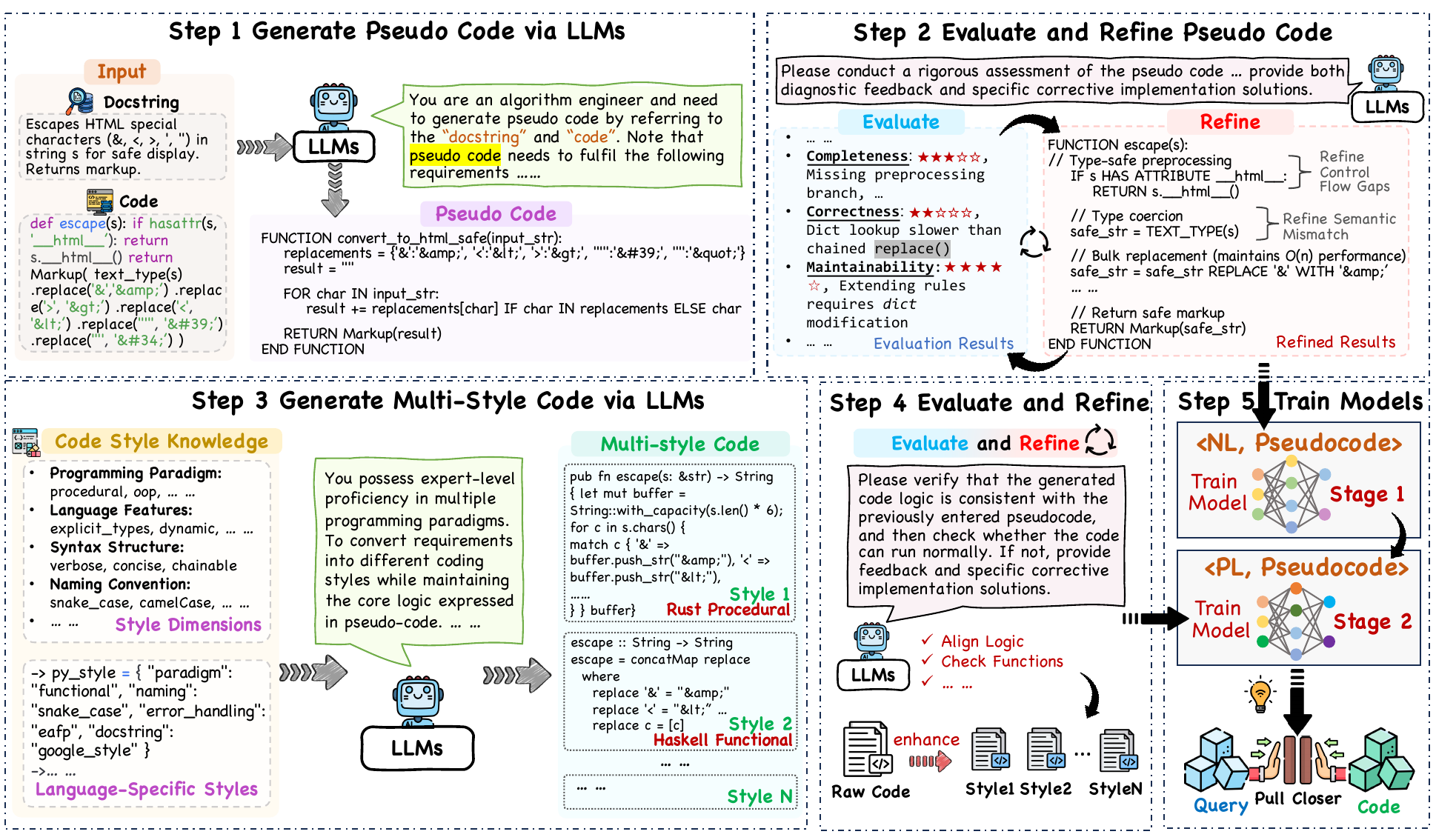}
    \caption{The framework of PseudoBridge, which comprises three core components: pseudo-code generation, logic-invariant code style enhancement, and model training. Step 1: Synthesize initial pseudo-code using LLMs. Step 2: Assess the quality of the generated pseudo-code and refine it. Step 3: Leverage the refined high-quality pseudo-code to produce syntactically diverse yet functionally equivalent code variants. Step 4: Evaluate the augmented code for logical correctness and quality, performing necessary refinement. Step 5: Utilize the generated pseudo-code, diversified code variants, and corresponding query to jointly train the target model.}
    \vspace{-0.2cm}
    \label{fig:PseudoBridge_overview}
\end{figure*}

\subsection{Pseudo-Code Generation}
\label{sec3.2}

Given the fundamental difference between NL and PL, an effective intermediate representation is essential to bridge the semantic gap. 
Unlike ASTs that focus on syntactic structures, pseudo-code preserves the logical flow of the PL while aligning closely with NL. 
However, existing code corpora typically lack pseudo-code annotations. 
To overcome this limitation, we propose a two-stage approach that leverages the LLM to synthesize pseudo-code from available <NL,PL> pairs. The first stage generates pseudo-code under well-defined constraints, and the second stage performs evaluation and refinement to ensure quality.

Specifically, given a corpus of NL and PL pairs $(Q_i, C_i)$, the first stage uses a carefully designed prompt framework $\mathbb{P}$ to guide the LLM in producing structurally consistent and logically clear pseudo-code $P_i$. These constraints include ``Naming Standardization'' for expanding abbreviations, ``Abstraction Standards'' for prioritizing key algorithmic steps, and ``Control Structures'' for enhancing logical clarity through standardized indentation. Detailed prompt templates are presented in part (a) of Figure~\ref{new_fig:prompt_a}.

To mitigate potential LLM hallucinations and ensure the quality of the synthesized data, the second stage incorporates an \textbf{Evaluate and Refine} mechanism.
It consists of three steps:

\begin{itemize}
    \item \textbf{Global Consistency Evaluation:} We employ LLMs to conduct a rigorous bidirectional validation against the source code and natural language queries to identify potential hard errors. 
    This validation mechanism encompasses three core dimensions. Firstly, \textbf{semantic consistency} ensures that the pseudo-code accurately aligns with the core operational intents described in the natural language query. Secondly, \textbf{logical equivalence} verifies whether the pseudo-code faithfully reproduces the control flow structures of the source code, such as loops and branching conditions. Finally, \textbf{hallucination checks} strictly detect fabricated steps that lack a basis in the source code. We immediately mark any output violating these constraints as a \textbf{``Hard Error''} and intercept it. Only samples that pass all checks proceed to the subsequent stage.
    
    \item \textbf{Fine-grained Quality Evaluation:} Following the preliminary validation of the generated pseudo-code, we implement a fine-grained multi-dimensional evaluation mechanism to quantify its quality. We ground this mechanism in Knuth's perspective~\cite{knuth1984literate} that pseudo-code acts as a bridge between humans and machines. We employ LLMs as automated evaluators to assess pseudo-code across five core dimensions, assigning scores ranging from 1 to 5. These dimensions include \textbf{correctness}, \textbf{readability}, \textbf{completeness}, \textbf{conciseness}, and \textbf{maintainability}. Part b of Figure~\ref{new_fig:prompt_a} in Appendix~\ref{appendix:a} details the definitions and scoring rubrics for each dimension. Unlike the binary check in the previous stage, this phase provides a granular quality profile. It aims to filter for high-quality pseudo-code that is not only logically accurate but also elegantly expressed and structurally clear. This process establishes a robust data foundation for the subsequent generation of multi-style code variants.
    \item \textbf{Feedback-Driven Iterative Refinement:} The refinement mechanism operates as a dynamic control loop integrating feedback from the preceding evaluation phases. 
    Specifically, the system triggers a targeted rewrite in two scenarios: (1) if the global verification module identifies a \textbf{``Hard Error''}, or (2) if the fine-grained quality assessment yields a score below 4.0 in any dimension. In each iteration, the LLM utilizes specific error feedback or diagnostics regarding low scores to guide the regeneration process. This \textbf{Evaluate and Refine} loop continues until the output satisfies all quality criteria. To balance computational costs and quality, we limit the maximum number of iterations to three. Experimental results indicate that this setting is sufficient for the majority of samples.
\end{itemize}

For detailed prompts and case studies, please refer to Appendix~\ref{appendix:a} Figure~\ref{new_fig:prompt_a} part (b) and Appendix~\ref{sec5.1.1} respectively.

\subsection{Multi-Style Code Generation}
\label{sec3.3}
Building on the synthesized high-quality pseudo-code, this module implements a logic-invariant strategy to explicitly decouple logical semantics from syntactic forms. As illustrated in Step 3 of Figure~\ref{fig:PseudoBridge_overview}, we establish a robust one-to-many mapping from a single pseudo-code representation to multiple functionally equivalent but stylistically diverse code implementations. This mechanism enables the model to learn invariant representations that are robust to stylistic variations.

\subsubsection{Multi-Dimensional Constraint Construction}
\begin{table}[htbp]
\caption{Six Core Dimensions of the Style Constraint Space}
\label{new_tab:mutli_style_def}
\centering
\small
\renewcommand{\arraystretch}{1.16} 

\begin{tabular}{|>{\centering\arraybackslash\bfseries}m{1.8cm}|m{7.7cm}|m{4.0cm}|}
\hline
\rowcolor[HTML]{EFEFEF} 
Dimension & \multicolumn{1}{c|}{\textbf{Description}} & \multicolumn{1}{c|}{\textbf{Examples}} \\ \hline

Programming \par Paradigm & 
Guides transitions across procedural, object-oriented, functional, and declarative paradigms, restructuring control flow and logical organization while preserving the underlying algorithm. & 
$\bullet$ Procedural: \texttt{for} loops \newline
$\bullet$ Functional: \texttt{map/reduce} \newline
$\bullet$ Recursive: Self-reference \\ \hline

Language \par Features & 
Varies the usage of type systems, such as explicit type declarations versus implicit type inference. It also adjusts the strictness of type enforcement to enhance adaptability to different language constraints.  & 
$\bullet$ Explicit: \texttt{int x = 5} \newline
$\bullet$ Implicit: \texttt{auto x = 5} \newline
$\bullet$ Lambda: \texttt{lambda a,b: a+b} \\ \hline

Syntactic \par Structures & 
Modulates conciseness by alternating between encapsulated chained calls and step-by-step execution, simulating coding styles of developers with varying expertise. & 
$\bullet$ Chained: \texttt{df.sort().head()} \newline
$\bullet$ Step-by-step: Multi-line calls \newline
$\bullet$ Ternary: \texttt{res = c ? a : b} \\ \hline

Naming \par Conventions & 
Enforces rotation among identifier naming schemes. This prevents the model from overfitting to specific naming patterns and directs focus toward the underlying data flow semantics. & 
$\bullet$ \texttt{snake\_case} \newline
$\bullet$ \texttt{camelCase} \newline
$\bullet$ \texttt{PascalCase} \\ \hline

Error \par Handling & 
Directs the model to switch between exception handling strategies, which introduces significant variations in execution paths.  & 
$\bullet$ LBYL: \texttt{if exists(file)} \newline
$\bullet$ EAFP: \texttt{try...catch} \\ \hline

Memory \par Management & 
Tailored for system-level languages, this approach varies resource management strategies such as manual allocation, smart pointers, and garbage collection to broaden the coverage of low-level logic. & 
$\bullet$ Manual: \texttt{malloc/free} \newline
$\bullet$ Automatic: Smart pointers \newline
$\bullet$ Explicit: \texttt{file.close()}  \\ \hline

\end{tabular}

\end{table}

To ensure functional equivalence alongside structural diversity, we design a style constraint space comprising six core dimensions. This framework mitigates the inherent bias of LLMs toward common patterns and defines explicit boundaries for style transformation. As detailed in Table~\ref{new_tab:mutli_style_def} and grounded in prior research~\cite{zhang2025function,van2001pep,oualline2002practical,reddy2000java}, these dimensions span from macro-level \textbf{Programming Paradigms} to micro-level \textbf{Syntactic Structures}. Additionally, we incorporate dimensions such as \textbf{Naming Conventions} and \textbf{Error Handling} to simulate diverse developer habits. Collectively, these orthogonal dimensions establish a robust search space for generating code that remains logically consistent yet syntactically distinct.

\subsubsection{Logic-Invariant Multi-style Generation}

Based on the multi-dimensional constraint space defined in the previous section, we utilize LLMs to transform pseudo-code into diverse implementations. This approach guides the model beyond conventional patterns to explicitly construct logical invariants. The process comprises two steps. First, during prompt construction, specific constraints are embedded into the template shown in Figure~\ref{new_fig:prompt_a} part (c) in Appendix~\ref{appendix:a}. This mandates adherence to style constraints while guiding the original logic. Second, during generation, the model executes significant structural adjustments. For instance, it refactors iterative logic into recursive operations or switches memory management strategies. This deep syntactic transformation compels the model to ignore superficial syntactic features and focus on the underlying logical essence.

\subsubsection{Evaluation and Iterative Refinement}

To ensure the quality of the generated code variants, we implement an LLM-based automated evaluation and iterative refinement strategy. This process comprises two core dimensions: logical consistency checking and functional simulation verification. 
First, the model compares the generated code with the original pseudo-code. It ensures a strict mapping of core algorithms, control flow, and data structures to maintain logical invariance. 
Subsequently, the model detects potential syntax errors and logical flaws by constructing representative input cases within a chain-of-thought process and simulating code execution. 
Based on the evaluation results, the system executes conditional repairs. For defective samples, the model performs targeted corrections on erroneous logic while strictly preserving the original style constraints, such as specific programming paradigms or naming conventions. In this paper, the maximum number of iterations for the \textbf{Evaluate and Refine} of multi-style code quality is set to three. Appendix~\ref{sec5.1.2} presents a detailed demonstration.

\subsection{Model Training}
\label{sec3.4}

This paper introduces PseudoBridge, a novel framework for code retrieval. 
The primary goal is to leverage LLMs to synthesize both pseudo-code and multi-style code variants, thereby enhancing the model’s capacity to align NL semantics with PL logic. 
This improved alignment translates into better code retrieval performance. 
The training process of the framework comprises two stages, as illustrated in Figure~\ref{fig:PseudoBridge_overview} Step 5.

In the first stage, LLMs are employed to synthesize a pseudo-code dataset. 
The model learns representations for both NL $Q$ and pseudo-code $P$, and is optimized to improve the alignment between them. 
The objective is to optimize the alignment between these two modalities by minimizing a similarity loss function $\mathcal{L}_{<Q, P>}$. The specific computation formula is as follows:

\begin{equation}
\begin{split}
\mathcal{L}_{<Q,P>} = & -\frac{1}{B} \sum_{i=1}^{|B|} 
\log \left( \frac{e^{\phi(q_i, p_i)}}{\textstyle e^{\phi(q_i, p_i)} + \sum_{p_j \in \mathcal{N}_{(p_i^-)}} e^{\phi(q_i, p_j)}} \right)
\end{split}
\end{equation}

where $B$ denotes the batch size, $q_i$ is the NL query of the $i$-th sample, $p_i$ is the corresponding pseudo-code, $\mathcal{N}_{(p_i^-)}$ contains pseudo-code negatives from other samples in the batch, and $\phi$ denotes the cosine similarity function.

In the second stage, based on the synthesized pseudo-code, LLMs generate code variants that are functionally equivalent but stylistically diverse. 
The model is further trained on these diverse code samples along with their corresponding pseudo-code, which strengthens the alignment between PL code representations $C$ and pseudo-code representations $P$. The loss function is formulated as:

\begin{equation}
\mathcal{L}_{<C,P>} = -\frac{1}{B} \sum_{i=1}^{|B|} \log \left( \frac{\sum_{c_{k}^{+} \in \mathcal{P}_{i}} e^{\phi(p_i, c_{k}^{+})}}{\sum_{c_{k}^{+} \in \mathcal{P}_{i}} e^{\phi(p_i, c_{k}^{+})} + \sum_{c_{j} \in \mathcal{N}_{(c_i^-)}} e^{\phi(p_i, c_{j})}} \right)
\end{equation}

Here, $\mathcal{P}_i$ denotes the set of positive code variants with stylistic diversity for the $i$-th sample, $c_{k}^{+}$ is the $k$-th positive code variant in $\mathcal{P}_i$, $\mathcal{N}_{(c_i^-)}$ represents the set of irrelevant code samples in the batch, and $c_j$ is the $j$-th irrelevant code sample.

\section{Experiments}

\subsection{Experimental Setup}

\subsubsection{Dataset}
\label{section:dataset}

Table~\ref{tab:dataset_info} summarizes the statistics for the training and evaluation datasets. CodeSearchNet~\cite{husain2019codesearchnet} is employed as the benchmark to assess retrieval performance. To construct a high-quality training set, the framework follows the CodeXGLUE~\cite{lu2021codexglue} filtering pipeline. Following initial processing, Stage 1 synthesizes 19,000 high-quality pseudo-code samples adhering to the established data distribution. In Stage 2, stratified sampling identifies 6,500 samples for logic-invariant, multi-style code augmentation. Each selected instance produces four functionally equivalent variants with distinct syntactic styles, yielding a final core training corpus of 26,000 enhanced code entries.  

For the testing phase, to simulate a real-world code retrieval scenario, we directly evaluate the model on the original CodeSearchNet~\cite{husain2019codesearchnet} test set. Additionally, to validate the zero-shot generalization performance, we include Solidity~\cite{chai2022cross} as well as C++ and C\# datasets from XLCoST~\cite{zhu2022xlcost} for evaluation. The details of the dataset are as follows:

\textbullet~\textbf{CodeSearchNet~\cite{husain2019codesearchnet}} corpus is a large-scale function-level dataset comprising code and corresponding documentation collected from open-source GitHub projects. It encompasses six PLs: Go, Java, JavaScript, PHP, Python, and Ruby. The dataset contains approximately 6 million code snippets, of which 2 million form valid <comment, code> pairs.

\textbullet~\textbf{XLCoST~\cite{zhu2022xlcost}} is collected from the GeeksForGeeks platform and comprises a series of programming problems along with their corresponding solutions, covering seven commonly-used programming languages. To evaluate the zero-shot cross-lingual generalization capability of PseudoBridge, this study selects its C++ and C$\#$ subsets for experimentation.

\textbullet~\textbf{Solidity~\cite{chai2022cross}}, a high-level programming language designed for smart contract development, is frequently adopted as an evaluation benchmark for cross-domain code retrieval tasks due to its domain-specific nature. We utilize this dataset to assess the zero-shot performance of PseudoBridge in retrieving code written in blockchain programming languages.

\begin{table}[htbp]
  \centering
  \setlength{\abovecaptionskip}{0.1cm} 
  \setlength{\belowcaptionskip}{-0.1cm}
  \setlength{\tabcolsep}{4pt} 
  \caption{Statistics of the datasets.}  
  \resizebox{0.74\linewidth}{!}{
    \begin{tabular}{c|c|c||c|c|c}
    \toprule
    \textbf{Language} & \textbf{For Training} & \textbf{For Testing} & \textbf{Language} & \textbf{For Training} & \textbf{For Testing} \\
    \midrule
    \textbf{Python} & 5,914 & 22,176 & \textbf{PHP} & 1,000 & 28,391 \\
    \textbf{Java} & 5,086 & 26,909 & \textbf{C++} & -     & 899 \\
    \textbf{JavaScript} & 5,000 & 6,483 & \textbf{C\#} & -     & 909 \\
    \textbf{Go} & 1,000 & 14,291 & \textbf{Solidity} & -     & 1,000 \\
    \textbf{Ruby} & 1,000 & 2,279 & ---     & ---     & --- \\
    \bottomrule
    \end{tabular}%
    }
    \vspace{-0.15cm}
  \label{tab:dataset_info}%
\end{table}%


\subsubsection{Metrics}
We adopt two widely recognized evaluation metrics, Mean Reciprocal Rank (MRR) and $Recall@k$ (with k = 1) to assess the performance of code retrieval in the test set. These metrics are commonly used in code search literature~\cite{chai2022cdcs,fan2024rapid,liang2025codebridge} for measuring retrieval effectiveness.

\textbf{MRR} quantifies the quality of ranked retrieval results by evaluating the reciprocal rank of the first relevant item. 
The Reciprocal Rank (RR) for a single query and the Mean Reciprocal Rank (MRR) over a query set are defined as:
\begin{equation}
    \mathrm{RR} = \frac{1}{\mathrm{rank}_i}, \;
    \mathrm{MRR} = \frac{1}{|Q|} \sum_{i=1}^{|Q|} \frac{1}{\mathrm{rank}_i}
\end{equation}
where $rank_i$ denotes the rank position (starting at 1) of the first correct result for the $i$-th query, $Q$ represents the query set and $|Q|$ is the total number of queries.

\textbf{Recall@k} measures the proportion of queries for which at least one relevant result appears within the $top-k$ ranked items. Its query-level and macro-averaged formulations are:

\begin{gather}
    \mathrm{Recall}@k^{(i)} = \frac{m_k^i}{M_i}, \; 
    \mathrm{Recall}@k = \frac{1}{N} \sum_{i=1}^{N} \mathrm{Recall}@k^{(i)}
\end{gather}
where $m_k^i$ is the number of relevant items in the top-$k$ results for the $i$-th query,
$M_i$ is the total relevant items for the $i$-th query,
$N$ is the number of queries.

\begin{table*}[htbp]
  \centering
  \setlength{\abovecaptionskip}{0.1cm} 
  \setlength{\belowcaptionskip}{-0.1cm}
  \centering
  \caption{Performance of PseudoBridge on code retrieval across different PLMs at MRR \textbf{(RQ1)}. \textit{Note: Row 1 presents the baseline performance of various PLMs on the code retrieval task. Row 2 shows the results after fine-tuning with PseudoBridge.}}
   \resizebox{0.94\linewidth}{!}{
    \begin{tabular}{r|llllll}
    \toprule
    \multicolumn{1}{c|}{\textbf{Methods}} & \textbf{Python} & \textbf{Java} & \textbf{JavaScript} & \textbf{Go} & \textbf{Ruby} & \textbf{PHP} \\
    \midrule
    \multicolumn{1}{l|}{\textbf{CDCS~\cite{chai2022cdcs}}} & 0.0380  & 0.0168  & 0.0120  & 0.0188  & 0.0182  & 0.0046  \\
    \textit{+ PseudoBridge} 
    & \textbf{0.8226}$^{\textcolor{blue}{\uparrow78.5\%}}$ 
    & \textbf{0.5890}$^{\textcolor{blue}{\uparrow57.2\%}}$ 
    & \textbf{0.5627}$^{\textcolor{blue}{\uparrow55.1\%}}$ 
    & \textbf{0.6346}$^{\textcolor{blue}{\uparrow61.6\%}}$ 
    & \textbf{0.5985}$^{\textcolor{blue}{\uparrow58.0\%}}$ 
    & \textbf{0.5819}$^{\textcolor{blue}{\uparrow57.7\%}}$ \\
    \midrule
    \multicolumn{1}{l|}{\textbf{CodeT5~\cite{wang2021codet5}}} & 0.0494  & 0.0670  & 0.0554  & 0.1106  & 0.1159  & 0.0473  \\
    \textit{+ PseudoBridge} 
    & \textbf{0.2011}$^{\textcolor{blue}{\uparrow15.2\%}}$ 
    & \textbf{0.1993}$^{\textcolor{blue}{\uparrow13.2\%}}$ 
    & \textbf{0.1852}$^{\textcolor{blue}{\uparrow13.0\%}}$
    & \textbf{0.1939}$^{\textcolor{blue}{\uparrow8.3\%}}$
    & \textbf{0.2676}$^{\textcolor{blue}{\uparrow15.2\%}}$
    & \textbf{0.1621}$^{\textcolor{blue}{\uparrow11.5\%}}$ \\
    \midrule
    \multicolumn{1}{l|}{\textbf{RoBERTa~\cite{liu2019roberta}}} & 0.0383  & 0.0173  & 0.0132  & 0.0188  & 0.0200  & 0.0146  \\
    \textit{+ PseudoBridge} 
    & \textbf{0.8191}$^{\textcolor{blue}{\uparrow78.1\%}}$ 
    & \textbf{0.5824}$^{\textcolor{blue}{\uparrow56.5\%}}$
    & \textbf{0.5559}$^{\textcolor{blue}{\uparrow54.3\%}}$
    & \textbf{0.6404}$^{\textcolor{blue}{\uparrow62.2\%}}$
    & \textbf{0.6082}$^{\textcolor{blue}{\uparrow58.8\%}}$
    & \textbf{0.5681}$^{\textcolor{blue}{\uparrow55.4\%}}$ \\
    \midrule
    \multicolumn{1}{l|}{\textbf{CodeBert~\cite{feng2020codebert}}} & 0.0052  & 0.0017  & 0.0034  & 0.0026  & 0.0068  & 0.0010  \\
    \textit{+ PseudoBridge} 
    & \textbf{0.8435}$^{\textcolor{blue}{\uparrow83.8\%}}$
    & \textbf{0.6455}$^{\textcolor{blue}{\uparrow64.4\%}}$
    & \textbf{0.6259}$^{\textcolor{blue}{\uparrow62.3\%}}$
    & \textbf{0.6862}$^{\textcolor{blue}{\uparrow68.4\%}}$
    & \textbf{0.6831}$^{\textcolor{blue}{\uparrow67.6\%}}$
    & \textbf{0.6146}$^{\textcolor{blue}{\uparrow61.4\%}}$ \\
    \midrule
    \multicolumn{1}{l|}{\textbf{GraphCodeBert~\cite{guo2020graphcodebert}}} & 0.0719  & 0.0643  & 0.0556  & 0.0742  & 0.1085  & 0.0298  \\
    \textit{+ PseudoBridge} 
    & \textbf{0.8721}$^{\textcolor{blue}{\uparrow80.0\%}}$
    & \textbf{0.6781}$^{\textcolor{blue}{\uparrow61.4\%}}$
    & \textbf{0.6656}$^{\textcolor{blue}{\uparrow61.0\%}}$ 
    & \textbf{0.7092}$^{\textcolor{blue}{\uparrow63.5\%}}$ 
    & \textbf{0.7203}$^{\textcolor{blue}{\uparrow61.2\%}}$ 
    & \textbf{0.6411}$^{\textcolor{blue}{\uparrow61.1\%}}$ \\
    \midrule
    \multicolumn{1}{l|}{\textbf{DistilBert~\cite{sanh2019distilbert}}} & 0.0796  & 0.0521  & 0.0498  & 0.0481  & 0.1112  & 0.0347  \\
    \textit{+ PseudoBridge} 
    & \textbf{0.7848}$^{\textcolor{blue}{\uparrow70.5\%}}$ 
    & \textbf{0.4974}$^{\textcolor{blue}{\uparrow44.5\%}}$ 
    & \textbf{0.4895}$^{\textcolor{blue}{\uparrow44.0\%}}$ 
    & \textbf{0.5368}$^{\textcolor{blue}{\uparrow48.9\%}}$ 
    & \textbf{0.5907}$^{\textcolor{blue}{\uparrow48.0\%}}$ 
    & \textbf{0.5106}$^{\textcolor{blue}{\uparrow47.6\%}}$ \\
    \midrule
    \multicolumn{1}{l|}{\textbf{UniXcoder~\cite{guo2022unixcoder}}} & 0.6238  & 0.4970  & 0.4805  & 0.5119  & 0.5446  & 0.4439  \\
    \textit{+ PseudoBridge} 
    & \textbf{0.8360}$^{\textcolor{blue}{\uparrow21.2\%}}$ 
    & \textbf{0.6762}$^{\textcolor{blue}{\uparrow17.9\%}}$ 
    & \textbf{0.6308}$^{\textcolor{blue}{\uparrow15.0\%}}$ 
    & \textbf{0.6589}$^{\textcolor{blue}{\uparrow14.7\%}}$ 
    & \textbf{0.6704}$^{\textcolor{blue}{\uparrow12.6\%}}$ 
    & \textbf{0.6392}$^{\textcolor{blue}{\uparrow19.5\%}}$ \\
    \midrule
    \multicolumn{1}{l|}{\textbf{SentenceBert~\cite{reimers2019sentencebert}}} & 0.6310  & 0.4998  & 0.5237  & 0.6363  & 0.6294  & 0.4867  \\
    \textit{+ PseudoBridge} 
    & \textbf{0.8507}$^{\textcolor{blue}{\uparrow22.0\%}}$ 
    & \textbf{0.6317}$^{\textcolor{blue}{\uparrow13.2\%}}$ 
    & \textbf{0.6116}$^{\textcolor{blue}{\uparrow8.8\%}}$ 
    & \textbf{0.6769}$^{\textcolor{blue}{\uparrow4.1\%}}$ 
    & \textbf{0.6949}$^{\textcolor{blue}{\uparrow6.6\%}}$ 
    & \textbf{0.6241}$^{\textcolor{blue}{\uparrow13.8\%}}$ \\
    \midrule
    \multicolumn{1}{l|}{\textbf{CoCoSoDa~\cite{shi2023cocosoda}}} & 0.7875  & 0.6336  & 0.6151  & 0.6747  & 0.6714  & 0.5568  \\
    \textit{+ PseudoBridge} 
    & \textbf{0.8370}$^{\textcolor{blue}{\uparrow5.0\%}}$ 
    & \textbf{0.6800}$^{\textcolor{blue}{\uparrow4.6\%}}$ 
    & \textbf{0.6460}$^{\textcolor{blue}{\uparrow3.1\%}}$ 
    & \textbf{0.6795}$^{\textcolor{blue}{\uparrow0.5\%}}$ 
    & \textbf{0.6886}$^{\textcolor{blue}{\uparrow1.7\%}}$ 
    & \textbf{0.6378}$^{\textcolor{blue}{\uparrow8.1\%}}$ \\
    \midrule
    \multicolumn{1}{l|}{\textbf{RAPID~\cite{fan2024rapid}}} & 0.7476  & 0.6144  & 0.5793  & 0.6131  & 0.6363  & 0.5467  \\
    \textit{+ PseudoBridge} 
    & \textbf{0.8168}$^{\textcolor{blue}{\uparrow7.0\%}}$ 
    & \textbf{0.6406}$^{\textcolor{blue}{\uparrow2.6\%}}$ 
    & \textbf{0.6066}$^{\textcolor{blue}{\uparrow2.7\%}}$ 
    & \textbf{0.6381}$^{\textcolor{blue}{\uparrow2.5\%}}$ 
    & \textbf{0.6498}$^{\textcolor{blue}{\uparrow1.4\%}}$ 
    & \textbf{0.5841}$^{\textcolor{blue}{\uparrow3.7\%}}$ \\
    \bottomrule
    \end{tabular}%
    }
  \label{tab:main_results_mrr}%
\end{table*}%

\begin{table*}[htbp]
  \centering
  \setlength{\abovecaptionskip}{0.1cm} 
  \setlength{\belowcaptionskip}{-0.1cm}
  \centering
  \caption{Performance of PseudoBridge on code retrieval across different PLMs at Recall@1 \textbf{(RQ1)}. \textit{Note: Row 1 presents the baseline performance of various PLMs on the code retrieval task. Row 2 shows the results after fine-tuning with PseudoBridge.}}
   \resizebox{0.94\linewidth}{!}{
    \begin{tabular}{r|llllll}
    \toprule
    \multicolumn{1}{c|}{\textbf{Methods}} & \textbf{Python} & \textbf{Java} & \textbf{JavaScript} & \textbf{Go} & \textbf{Ruby} & \textbf{PHP} \\
    \midrule
    \multicolumn{1}{l|}{\textbf{CDCS~\cite{chai2022cdcs}}} & 0.0292  & 0.0103  & 0.0057  & 0.0096  & 0.0083  & 0.0023  \\
    \textit{+ PseudoBridge} 
    & \textbf{0.7709}$^{\textcolor{blue}{\uparrow74.2\%}}$ 
    & \textbf{0.4985}$^{\textcolor{blue}{\uparrow48.8\%}}$  
    & \textbf{0.4796}$^{\textcolor{blue}{\uparrow47.4\%}}$  
    & \textbf{0.5532}$^{\textcolor{blue}{\uparrow54.4\%}}$  
    & \textbf{0.5046}$^{\textcolor{blue}{\uparrow49.6\%}}$  
    & \textbf{0.4961}$^{\textcolor{blue}{\uparrow49.4\%}}$  \\
    \midrule
    \multicolumn{1}{l|}{\textbf{CodeT5~\cite{wang2021codet5}}} & 0.0362  & 0.0417  & 0.0299  & 0.0829  & 0.0702  & 0.0297  \\
    \textit{+ PseudoBridge} 
    & \textbf{0.1586}$^{\textcolor{blue}{\uparrow12.2\%}}$  
    & \textbf{0.1403}$^{\textcolor{blue}{\uparrow9.9\%}}$  
    & \textbf{0.1248}$^{\textcolor{blue}{\uparrow9.5\%}}$  
    & \textbf{0.1456}$^{\textcolor{blue}{\uparrow6.3\%}}$  
    & \textbf{0.1887}$^{\textcolor{blue}{\uparrow11.9\%}}$  
    & \textbf{0.1157}$^{\textcolor{blue}{\uparrow8.6\%}}$  
    \\
    \midrule
    \multicolumn{1}{l|}{\textbf{RoBERTa~\cite{liu2019roberta}}} & 0.0388  & 0.0157  & 0.0123  & 0.0109  & 0.0192  & 0.0110  \\
    \textit{+ PseudoBridge} 
    & \textbf{0.7666}$^{\textcolor{blue}{\uparrow72.9\%}}$  
    & \textbf{0.4893}$^{\textcolor{blue}{\uparrow47.4\%}}$  
    & \textbf{0.4700}$^{\textcolor{blue}{\uparrow45.8\%}}$  
    & \textbf{0.5579}$^{\textcolor{blue}{\uparrow54.7\%}}$  
    & \textbf{0.5186}$^{\textcolor{blue}{\uparrow49.9\%}}$  
    & \textbf{0.4808}$^{\textcolor{blue}{\uparrow47.0\%}}$  \\
    \midrule
    \multicolumn{1}{l|}{\textbf{CodeBert~\cite{feng2020codebert}}} & 0.0029  & 0.0006  & 0.0011  & 0.0008  & 0.0013  & 0.0002  \\
    \textit{+ PseudoBridge} 
    & \textbf{0.7946}$^{\textcolor{blue}{\uparrow79.2\%}}$  
    & \textbf{0.5583}$^{\textcolor{blue}{\uparrow55.8\%}}$  
    & \textbf{0.5439}$^{\textcolor{blue}{\uparrow54.3\%}}$  
    & \textbf{0.6152}$^{\textcolor{blue}{\uparrow61.4\%}}$  
    & \textbf{0.5919}$^{\textcolor{blue}{\uparrow59.1\%}}$  
    & \textbf{0.5346}$^{\textcolor{blue}{\uparrow53.4\%}}$  \\
    \midrule
    \multicolumn{1}{l|}{\textbf{GraphCodeBert~\cite{guo2020graphcodebert}}} & 0.0487  & 0.0395  & 0.0295  & 0.0442  & 0.0693  & 0.0172  \\
    \textit{+ PseudoBridge} 
    & \textbf{0.8277}$^{\textcolor{blue}{\uparrow77.9\%}}$  
    & \textbf{0.5949}$^{\textcolor{blue}{\uparrow55.5\%}}$  
    & \textbf{0.5865}$^{\textcolor{blue}{\uparrow55.7\%}}$  
    & \textbf{0.6417}$^{\textcolor{blue}{\uparrow59.8\%}}$  
    & \textbf{0.6384}$^{\textcolor{blue}{\uparrow56.9\%}}$  
    & \textbf{0.5635}$^{\textcolor{blue}{\uparrow54.6\%}}$  \\
    \midrule
    \multicolumn{1}{l|}{\textbf{DistilBert~\cite{sanh2019distilbert}}} & 0.0600  & 0.0321  & 0.0292  & 0.0293  & 0.0706  & 0.0223  \\
    \textit{+ PseudoBridge} 
    & \textbf{0.7257}$^{\textcolor{blue}{\uparrow66.6\%}}$  
    & \textbf{0.4019}$^{\textcolor{blue}{\uparrow37.0\%}}$  
    & \textbf{0.4001}$^{\textcolor{blue}{\uparrow37.1\%}}$  
    & \textbf{0.4478}$^{\textcolor{blue}{\uparrow41.8\%}}$  
    & \textbf{0.4958}$^{\textcolor{blue}{\uparrow42.5\%}}$  
    & \textbf{0.4205}$^{\textcolor{blue}{\uparrow39.8\%}}$  \\
    \midrule
    \multicolumn{1}{l|}{\textbf{UniXcoder~\cite{guo2022unixcoder}}} & 0.5600  & 0.3976  & 0.3873  & 0.4199  & 0.4506  & 0.3531  \\
    \textit{+ PseudoBridge} 
    & \textbf{0.7886}$^{\textcolor{blue}{\uparrow22.9\%}}$  
    & \textbf{0.5976}$^{\textcolor{blue}{\uparrow20.0\%}}$  
    & \textbf{0.5521}$^{\textcolor{blue}{\uparrow16.5\%}}$  
    & \textbf{0.5805}$^{\textcolor{blue}{\uparrow16.1\%}}$  
    & \textbf{0.5814}$^{\textcolor{blue}{\uparrow13.1\%}}$  
    & \textbf{0.5600}$^{\textcolor{blue}{\uparrow20.7\%}}$  \\
    \midrule
    \multicolumn{1}{l|}{\textbf{SentenceBert~\cite{reimers2019sentencebert}}} & 0.5556  & 0.3896  & 0.4211  & 0.5503  & 0.5195  & 0.3905  \\
    \textit{+ PseudoBridge} 
    & \textbf{0.8028}$^{\textcolor{blue}{\uparrow24.7\%}}$  
    & \textbf{0.5411}$^{\textcolor{blue}{\uparrow15.2\%}}$  
    & \textbf{0.5275}$^{\textcolor{blue}{\uparrow10.6\%}}$  
    & \textbf{0.6006}$^{\textcolor{blue}{\uparrow5.0\%}}$  
    & \textbf{0.6042}$^{\textcolor{blue}{\uparrow8.5\%}}$  
    & \textbf{0.5414}$^{\textcolor{blue}{\uparrow15.1\%}}$  \\
    \midrule
    \multicolumn{1}{l|}{\textbf{CoCoSoDa~\cite{shi2023cocosoda}}} & 0.7306  & 0.5394  & 0.5249  & 0.6028  & 0.5753  & 0.4637  \\
    \textit{+ PseudoBridge} 
    & \textbf{0.7873}$^{\textcolor{blue}{\uparrow5.7\%}}$  
    & \textbf{0.5998}$^{\textcolor{blue}{\uparrow6.0\%}}$  
    & \textbf{0.5666}$^{\textcolor{blue}{\uparrow4.2\%}}$  
    & \textbf{0.6104}$^{\textcolor{blue}{\uparrow0.8\%}}$  
    & \textbf{0.5954}$^{\textcolor{blue}{\uparrow2.0\%}}$  
    & \textbf{0.5568}$^{\textcolor{blue}{\uparrow9.3\%}}$  
    \\
    \midrule
    \multicolumn{1}{l|}{\textbf{RAPID~\cite{fan2024rapid}}} & 0.6884  & 0.5205  & 0.4894  & 0.5287  & 0.5406  & 0.4590  \\
    \textit{+ PseudoBridge} 
    & \textbf{0.7644}$^{\textcolor{blue}{\uparrow7.6\%}}$  
    & \textbf{0.5516}$^{\textcolor{blue}{\uparrow3.1\%}}$  
    & \textbf{0.5198}$^{\textcolor{blue}{\uparrow3.0\%}}$  
    & \textbf{0.5572}$^{\textcolor{blue}{\uparrow2.9\%}}$  
    & \textbf{0.5546}$^{\textcolor{blue}{\uparrow1.4\%}}$  
    & \textbf{0.4968}$^{\textcolor{blue}{\uparrow3.8\%}}$  \\
    \bottomrule
    \end{tabular}%
    }
    \vspace{-0.15cm}
  \label{tab:main_results_recall}%
\end{table*}%

\subsection{Baselines}

Current pre-trained models for code retrieval generally fall into three primary categories based on their architectures: Encoder-only Models, Unified and Generative Models, and Specific Retrieval Frameworks. 
As PseudoBridge employs pseudo-code as an intermediate bridge to align NL semantics with PL logic, it is essential to assess its universality. To comprehensively evaluate this effectiveness, we select representative models from each of the three categories as baselines. 
These models span diverse technical paradigms, ranging from general natural language processing to specialized code intelligence. This selection allows us to verify that PseudoBridge delivers consistent performance improvements across different underlying architectures.

\subsubsection{Encoder-only Models}
The first category comprises encoder-only models, which constitute the dominant architecture for code retrieval. To evaluate the adaptability of our framework across diverse domains, we categorize these models into general natural language models and code-specific models. For the general natural language models, we select three representative works.

\begin{itemize}
    \item \textbf{RoBERTa}~\cite{liu2019roberta} is built on a multi-layer Transformer encoder and is pre-trained using masked language modeling on natural language and source code corpora. 
    \item \textbf{DistilBERT}~\cite{sanh2019distilbert} is a lightweight version of BERT created via distillation. 
    It maintains performance comparable to the original model while significantly reducing parameter size and improving inference speed.
    \item \textbf{SentenceBERT}~\cite{reimers2019sentencebert} enhances BERT by using a siamese network structure. It generates semantically comparable sentence embeddings, which significantly improves the efficiency of similarity computation.
\end{itemize}

For code-specific models, we select two classic bimodal encoders.
\begin{itemize}
    \item \textbf{CodeBERT}~\cite{feng2020codebert} is a pre-trained model for both code and natural language. It effectively learns bimodal representations using replaced token detection and a hybrid training strategy. 
    \item \textbf{GraphCodeBERT}~\cite{guo2020graphcodebert} builds upon this by incorporating structural information, such as data flow. This allows it to more accurately capture the intrinsic logic and dependencies of code. 
\end{itemize}

\subsubsection{Unified and Generative Models}

The second category covers unified and generative models, aiming to verify the compatibility of PseudoBridge with architectures beyond encoder-only models. 
By experimenting with these structures, we demonstrate that our framework is not only suitable for discriminative retrieval tasks but also effectively enhances the performance of generative and unified models. 
This highlights the broad architectural adaptability of our approach. We choose two representative models in this category.

\begin{itemize}
    \item \textbf{UniXcoder}~\cite{guo2022unixcoder} is a unified multimodal pre-trained model. It takes code summaries and simplified ASTs as input, incorporating multiple training objectives such as language modeling, denoising, and contrastive learning. 
    \item \textbf{CodeT5}~\cite{wang2021codet5} is a sequence-to-sequence pre-trained model based on an encoder-decoder architecture. It utilizes identifier-aware pre-training tasks to improve the understanding and recovery of code semantics.
\end{itemize}

\subsubsection{Specific Retrieval Models}
The third category consists of  high-performance frameworks specifically optimized for code retrieval. We include these to determine whether PseudoBridge offers incremental gains over existing advanced methods. By selecting three representative approaches, we demonstrate that our framework further enhances robustness and accuracy, even when applied to highly optimized baselines.

\begin{itemize}
    \item \textbf{CDCS}~\cite{chai2022cdcs} employs a meta-learning approach to effectively transfer knowledge from general pre-trained models to domain-specific languages. 
    \item \textbf{CoCoSoDa}~\cite{shi2023cocosoda} optimizes the alignment of multimodal representations through contrastive learning equipped with soft data augmentation and momentum-based negative sampling. 
    \item \textbf{RAPID}~\cite{fan2024rapid}, evaluated here in its UniXcoder-based version, targets zero-shot domain adaptation by leveraging pseudo-labeled synthetic data and mixed sampling strategies.
\end{itemize}

\subsection{Implementation Details}

\subsubsection{Data Synthesis and Quality Assurance Settings}

PseudoBridge relies on advanced LLMs to synthesize pseudo-code and generate multi-style code variants. In our experiments, we utilize \textit{\textbf{GPT-4o}} (\textit{version: gpt-4o-2024-08-06}) as the core generation engine. To balance generation stability and diversity, we set the temperature parameter to 0.6 for the pseudo-code generation phase and adjust it to 0.8 for the multi-style code generation phase. The detailed hyperparameters are summarized in Table~\ref{tab:llm_parameters} of Appendix~\ref{appendix_llm_parameters}.
For automated quality assessment, we employ \textit{\textbf{DeepSeek-R1-Distill-Llama-70B}}~\cite{guo2025deepseekr1} as an automatic evaluator. This evaluation process operates in two stages. First, the evaluator diagnoses the semantic consistency and logical correctness of the synthesized pseudo-code compared to the natural language and source code. Second, it verifies the execution and functional correctness of the generated multi-style code variants.

To verify the reliability of the automated \textbf{``Evaluate and Refine''} mechanism, we perform a rigorous manual audit following standard dataset construction protocols.

\begin{itemize}
    \item \textbf{Statistical Sampling.} We use the Cochran formula~\cite{cochran1977sampling} to determine the sample size and ensure that the manual audit represents the total population. In Stage 1, we select 400 samples from 19,000 pseudo-code entries to achieve a 95\% confidence level with a 4.8\% margin of error. In Stage 2, we apply the same statistical standard to the 26,000 multi-style code variants generated from 6,500 seeds. We randomly select 100 sample groups where each group includes all four style variants for a total of 400 code items. This strategy ensures that the samples are representative and accurately reflect the overall quality of the synthetic corpus.
    \item \textbf{Audit Protocol.} To ensure objectivity and rigor, two independent researchers with software engineering expertise and no project involvement conduct a double blind review and audit. A sample is deemed valid only upon mutual consensus. The audit evaluates \textbf{pseudo-code quality} by assessing the accurate restoration of core algorithmic logic across five predefined dimensions and verifies \textbf{code variant validity} by ensuring functional equivalence to the original code under strict style constraints.
    \item \textbf{Consistency Analysis.} We use the Cohen's Kappa coefficient to measure the inter-rater reliability between the automated evaluator and human experts. In Stage 1, the agreement rate is 91.3\% with a Kappa of 0.82. In Stage 2, the agreement rate is 88.3\% with a Kappa of 0.79.
\end{itemize}

According to established statistical benchmarks~\cite{landis1977measurement}, Kappa values between 0.61 and 0.80 indicate substantial agreement. Our results demonstrate that the automated quality control process in PseudoBridge aligns closely with human expert judgment. This consistency confirms that the proposed mechanism can effectively replace high-cost manual intervention, ensuring the reliability and scalability of large-scale dataset construction.

\subsubsection{Model Tuning Settings}

We implement the fine-tuning of retrieval models using the PyTorch 2.6.0 framework and the HuggingFace Transformers library within a Python 3.10.16 environment. 
To evaluate the generalization of the PseudoBridge framework, we conduct experiments on ten pre-trained models with diverse architectures and parameter scales, such as CodeBERT, UniXcoder, and CoCoSoDa. We execute all training tasks in parallel on two NVIDIA A100-SXM4 GPUs (80GB). 
For the two-stage fine-tuning, we configure the hyperparameters as follows: a batch size of 48, a context window length of 2,048 tokens, and a learning rate of 5e-5. 
We use the AdamW optimizer coupled with a linear learning rate scheduler and train the models for 3 epochs. Additionally, we set the temperature parameter in the contrastive learning loss function to 0.05. 
Taking CodeBERT as an example, we conduct the training on two A100 GPUs. In the first fine-tuning stage, the process takes approximately 0.3 GPU hours, with a peak memory usage of 73,003 MB per card. In the second fine-tuning stage, the process takes about 0.7 GPU hours, and the peak memory usage reaches 76,320 MB per card. The detailed training hyperparameters are summarized in Table~\ref{tab:finetuning_parameters} of Appendix~\ref{appendix_finetuning_parameters}.

\subsubsection{Evaluation Settings}
To ensure efficiency and consistency in evaluation, we formulate the code retrieval task as a dense retrieval problem. 
Specifically, the models encode natural language queries and code snippets into high-dimensional semantic vectors. 
We adopt a batch processing strategy for embedding computation to maximize inference throughput, with a uniform batch size of 64. 
For specific architectures such as UniXcoder, we adjust the maximum sequence length to 1,024 to fully accommodate long code contexts. 
Given a test set with $N$ samples, we compute a pairwise cosine similarity matrix of dimension $N \times N$ between all query vectors and candidate code vectors. 
For each query, we rank all $N$ candidates in descending order based on similarity scores, treating the unique paired code snippet as the ground truth. 
We employ Recall@1 and Mean Reciprocal Rank (MRR) as the metrics for quantitative evaluation.

\subsection{Evaluations}

\subsubsection{\textbf{RQ1: How effective is PseudoBridge in bridging the gap between natural language semantics and programming language logic?}}
To verify the core efficacy of the PseudoBridge framework in improving the alignment between NL semantics and PL logic, we conduct comprehensive code retrieval experiments across six mainstream PLs: Python, Java, JavaScript, Go, Ruby, and PHP. This experiment aims to determine whether PseudoBridge consistently enhances code retrieval performance across PLMs with diverse architectures and capabilities. Consequently, we seek to demonstrate the universal effectiveness of the \textbf{pseudo-code as an intermediate bridge} mechanism.

\textbf{Setup.}
We apply PseudoBridge to ten representative baseline PLMs. We fine-tune all models on the CodeSearchNet dataset and evaluate them using MRR and Recall@1 metrics. To analyze the performance gains in detail, we categorize the baselines into two groups based on their architectural paradigms and initial retrieval performance:

\begin{itemize}
    \item \textbf{Low-Performance Models:} These primarily include standard Encoder-only architectures (e.g., RoBERTa, CodeBERT, GraphCodeBERT, DistilBERT). While possessing basic code representation capabilities, they exhibit relatively low baseline metrics in complex code retrieval tasks.
    \item \textbf{High-Performance Models:} These encompass unified and generative architectures (e.g., UniXcoder) and specialized retrieval frameworks (e.g., CoCoSoDa, SentenceBERT). Leveraging specialized pre-training objectives or contrastive learning strategies, these models achieve robust semantic understanding and strong performance.
\end{itemize}

\textbf{Results and Analysis.}
The experimental results are presented in Table~\ref{tab:main_results_mrr} (MRR) and Table~\ref{tab:main_results_recall} (Recall@1). Through analysis, we draw the following key conclusions:

\begin{itemize}
    \item \textbf{Fundamental improvement in low-performance models:} Experimental results indicate that PseudoBridge delivers transformative improvements for underperforming models that struggle to capture deep logical semantics. By introducing pseudo-code as an intermediate modality, PseudoBridge bridges this gap effectively. For instance, CodeBERT's Recall@1 on the Python dataset surges from a negligible 0.0029 to 0.7946, while GraphCodeBERT's MRR on Ruby increases from 0.1085 to 0.7203.
    These results demonstrate that the logical alignment mechanism fundamentally enhances the capabilities of low-performance models. It enables a performance leap, transforming them from essentially unusable to highly effective.
    \item \textbf{Continuous gains in high-performance models:} PseudoBridge provides consistent and robust performance gains for high-performance models (e.g., UniXcoder, RAPID) that already integrate ASTs, data flow information, or momentum contrastive learning. 
    This demonstrates its orthogonality to existing advanced techniques. For instance, on the Go dataset, the Recall@1 of UniXcoder increases from 0.4199 to 0.5805. 
    Similarly, the MRR of SentenceBERT on Python improves from 0.6310 to 0.8507. 
    These results confirm that even when models possess strong structural awareness, the explicit logic enhancement strategy of PseudoBridge helps them break through existing semantic understanding bottlenecks. It enables further improvements in retrieval accuracy.
\end{itemize}

\begin{tcolorbox}[
    colback=gray!5!white,
    colframe=gray!95!black,
    fonttitle=\bfseries,
    arc=2.8pt,
    boxrule=0.9pt,
    enhanced,
    boxsep=1.5pt,              
    left=1.5pt, right=1.5pt,     
    top=1.5pt, bottom=1.5pt,     
    before upper={\setstretch{0.9}}
]
\textbf{Conclusion 1}: PseudoBridge aligns NL semantics with PL logic via pseudo-code and style augmentation. The performance gains follow a pattern of substantial transformation for weaker models and supplementary enhancement for stronger ones. Specifically, the bridging effect is most pronounced for models with limited initial capabilities, while providing critical refinement for advanced models, thereby achieving consistent improvements across all baselines.
\end{tcolorbox}

\begin{table*}[htbp]
  \centering
  \setlength{\abovecaptionskip}{0.1cm} 
  \setlength{\belowcaptionskip}{-0.1cm}
  \setlength{\tabcolsep}{4pt} 
  \caption{Impact of pseudo code proportions on code retrieval performance of PLMs \textbf{(RQ2)}. \textit{Note: Python and Ruby CodeSearchNet~\cite{husain2019codesearchnet}. Bold numbers indicate the best performance metrics.}}
  \resizebox{\linewidth}{!}{ 
    \begin{tabular}{c|l|cc|cc|cc|cc|cc}
    \toprule
    \multirow{2}[3]{*}{\textbf{Language}} & \multicolumn{1}{c|}{\multirow{2}[3]{*}{\textbf{Methods}}} & \multicolumn{2}{c|}{\textbf{20\%}} & \multicolumn{2}{c|}{\textbf{40\%}} & \multicolumn{2}{c|}{\textbf{60\%}} & \multicolumn{2}{c|}{\textbf{80\%}} & \multicolumn{2}{c}{\textbf{100\%}} \\
\cmidrule{3-12}          &       & MRR   & Recall@1 & MRR   & Recall@1 & MRR   & Recall@1 & MRR   & Recall@1 & MRR   & Recall@1 \\
\midrule
    \multirow{6}[1]{*}{Python} & CodeBert~\cite{feng2020codebert} & 0.8386  & 0.7901  & 0.8377  & 0.7865  & 0.8359  & 0.7858  & \textbf{0.8439 } & \textbf{0.7960 } & 0.8435  & 0.7946  \\
          & GraphCodeBert~\cite{guo2020graphcodebert} & 0.8619  & 0.8156  & 0.8681  & 0.8244  & 0.8708  & 0.8261  & 0.8570  & 0.8108  & \textbf{0.8721 } & \textbf{0.8277 } \\
          & DistilBert~\cite{sanh2019distilbert} & 0.7599  & 0.6990  & 0.7772  & 0.7182  & 0.7803  & 0.7217  & 0.7818  & 0.7236  & \textbf{0.7848 } & \textbf{0.7257 } \\
          \cmidrule{2-12}
          & UniXcoder~\cite{guo2022unixcoder} & 0.8352  & 0.7869  & 0.8308  & 0.7807  & 0.8328  & 0.7832  & \textbf{0.8369 } & \textbf{0.7887 } & 0.8360  & 0.7886  \\
          & SentenceBert~\cite{reimers2019sentencebert} & 0.8481  & 0.8004  & \textbf{0.8534 } & \textbf{0.8061 } & 0.8445  & 0.7953  & 0.8523  & 0.8055  & 0.8507  & 0.8028  \\
          & CoCoSoDa~\cite{shi2023cocosoda} & 0.8350  & 0.7858  & \textbf{0.8381 } & \textbf{0.7903 } & 0.8360  & 0.7868  & 0.8377  & 0.7890  & 0.8370  & 0.7873  \\
    \midrule
    \multirow{6}[2]{*}{Ruby} & CodeBert~\cite{feng2020codebert} & 0.6779  & 0.5858  & 0.6844  & 0.5946  & \textbf{0.6845 } & \textbf{0.5950 } & 0.6826  & 0.5915  & 0.6831  & 0.5919  \\
          & GraphCodeBert~\cite{guo2020graphcodebert} & 0.7298  & 0.6463  & \textbf{0.7303 } &  \textbf{0.6476 }  & 0.7291  & 0.6463  & 0.7195  & 0.6340  & 0.7203  & 0.6384 \\
          & DistilBert~\cite{sanh2019distilbert} & 0.5754  & 0.4778  & 0.5810  & 0.4822  & 0.5877  & 0.4927  & 0.5886  & 0.4962  & \textbf{0.5907 } & \textbf{0.4958 } \\
          \cmidrule{2-12}
          & UniXcoder~\cite{guo2022unixcoder} & 0.6699  & 0.5813  & 0.6697  & 0.5792  & 0.6651  & 0.5752  & \textbf{0.6748 } & \textbf{0.5875 } & 0.6704  & 0.5813  \\
          & SentenceBert~\cite{reimers2019sentencebert} & 0.6945  & 0.6042  & 0.6936  & \textbf{0.6055 } & 0.6902  & 0.5989  & 0.6939  & 0.6024  & \textbf{0.6949 } & 0.6042  \\
          & CoCoSoDa~\cite{shi2023cocosoda} & 0.6884  & 0.5954  & 0.6888  & 0.5954 & \textbf{0.6895 }  & 0.5993  & 0.6828  & 0.5862  & 0.6886  & \textbf{0.5954 }  \\
    \bottomrule
    \end{tabular}%
    }
  \label{tab:pseudocode_combined}%
\end{table*}%

\subsubsection{\textbf{RQ2: How does the proportion of pseudo-code affect PseudoBridge's alignment between NL semantics and PL logic?}}
We perform a sensitivity analysis to assess the impact of pseudo-code data volume on PseudoBridge. By evaluating mixing ratios from 20\% to 100\%, we aim to quantify the efficacy of the intermediate modality in bridging the logic and semantic gap between NL and PL. Furthermore, we analyze the robustness of models with diverse baseline capabilities to varying degrees of pseudo-code augmentation.

\textbf{Setup.}
We construct five training sets with varying proportions of pseudo-code, then train and evaluate the models under strictly controlled experimental conditions. 
To analyze the relationship between base model capability and pseudo-code dependency, we select code models with distinct performance levels. 
These include low-performance models (CodeBERT, GraphCodeBERT, DistilBERT) and those with high-performance models (UniXcoder, SentenceBERT, CoCoSoDa). We test all models on the Python and Ruby datasets from CodeSearchNet, focusing on two key aspects: (1) the performance trends relative to changes in pseudo-code ratios; and (2) the differential requirements of diverse model architectures for semantic alignment assisted by pseudo-code.

\textbf{Results and Analysis.}The experimental results are presented in Table~\ref{tab:pseudocode_combined}. The data reveals a significant correlation between the models' foundational capabilities and the required proportion of pseudo-code.

\begin{itemize}
    \item \textbf{Low-performance models favor high-ratio data augmentation:} On the data-rich Python dataset, models with weaker baselines generally require high ratios of pseudo-code to achieve peak performance. Specifically, GraphCodeBERT and DistilBERT achieve their best MRR at a 100\% pseudo-code ratio, scoring 0.8721 and 0.7848, respectively. CodeBERT attains its optimal result of 0.8439 at an 80\% ratio. This trend suggests that these models, lacking intrinsic logical alignment mechanisms, rely heavily on pseudo-code as an explicit semantic bridge. Notably, the pure-text DistilBERT consistently requires 100\% augmentation across datasets, reinforcing the critical dependency of weaker architectures on the intermediate modality.
    \item \textbf{Low-ratio saturation and marginal benefits for high-performance models:} In contrast, high-performance models demonstrate a distinct ``early saturation'' pattern on the Python task. Specifically, SentenceBERT and CoCoSoDa achieve peak MRR scores of 0.8534 and 0.8381 with a pseudo-code ratio of only 40\%. Beyond this threshold, performance stabilizes or slightly declines. This occurs because these models establish strong cross-modal representation capabilities during pre-training via techniques such as Siamese Networks or Momentum Contrast. Consequently, a low ratio of pseudo-code provides sufficient logical cues to activate their alignment capabilities. Excessive ratios do not yield significant marginal benefits and may introduce redundant information.
\end{itemize}

\begin{tcolorbox}[
    colback=gray!5!white,
    colframe=gray!95!black,
    fonttitle=\bfseries,
    arc=2.8pt,
    boxrule=0.9pt,
    enhanced,
    boxsep=1.5pt,              
    left=1.5pt, right=1.5pt,     
    top=1.5pt, bottom=1.5pt,     
    before upper={\setstretch{0.9}}
]
\textbf{Conclusion 2}: Different amounts of pseudo-code have different effects on model enhancement for models with different performance levels. Specifically, low-performance models require high pseudo-code ratios to compensate for weak cross-modal representations. In contrast, high-performance models achieve early saturation with minimal data, yielding diminishing returns at higher ratios.
\end{tcolorbox}

\begin{table*}[htbp]
  \centering
    \setlength{\abovecaptionskip}{0.1cm} 
    \setlength{\belowcaptionskip}{-0.1cm}
  \setlength{\tabcolsep}{4pt} 
  \caption{Impact of code style enhancement quantity on code retrieval performance of PLMs \textbf{(RQ3)}. \textit{Note: Python and Ruby CodeSearchNet~\cite{husain2019codesearchnet}. Bold numbers indicate the best performance metrics.}}
  \resizebox{0.95\linewidth}{!}{ 
    \begin{tabular}{c|l|cc|cc|cc|cc}
    \toprule
    \multirow{2}[3]{*}{\textbf{Language}} & \multicolumn{1}{c|}{\multirow{2}[3]{*}{\textbf{Methods}}} & \multicolumn{2}{c|}{\textbf{1}} & \multicolumn{2}{c|}{\textbf{2}} & \multicolumn{2}{c|}{\textbf{3}} & \multicolumn{2}{c}{\textbf{4}} \\
\cmidrule{3-10}          &       & MRR   & Recall@1 & MRR   & Recall@1 & MRR   & Recall@1 & MRR   & Recall@1 \\
\midrule
    \multirow{6}[1]{*}{Python} & CodeBert~\cite{feng2020codebert} & 0.8377  & 0.7865  & 0.8359  & 0.7858  & \textbf{0.8439 } & \textbf{0.7960 } & 0.8435  & 0.7946  \\
          & GraphCodeBert~\cite{guo2020graphcodebert} & 0.8681  & 0.8244  & 0.8708  & 0.8261  & 0.8719  & 0.8270  & \textbf{0.8721 } & \textbf{0.8277 } \\
          & DistilBert~\cite{sanh2019distilbert} & 0.7772  & 0.7182  & 0.7803  & 0.7217  & 0.7818  & 0.7236  & \textbf{0.7848 } & \textbf{0.7257 } \\
          \cmidrule{2-10}
          & UniXcoder~\cite{guo2022unixcoder} & 0.8308  & 0.7807  & 0.8328  & 0.7832  & \textbf{0.8369 } & \textbf{0.7887 } & 0.8360  & 0.7886  \\
          & SentenceBert~\cite{reimers2019sentencebert} & \textbf{0.8534 } & \textbf{0.8061 } & 0.8445  & 0.7953  & 0.8523  & 0.8055  & 0.8507  & 0.8028  \\
          & CoCoSoDa~\cite{shi2023cocosoda} & \textbf{0.8381 } & \textbf{0.7903 } & 0.8360  & 0.7868  & 0.8377  & 0.7890  & 0.8370  & 0.7873  \\
    \midrule
    \multirow{6}[2]{*}{Ruby} & CodeBert~\cite{feng2020codebert} & 0.6844  & 0.5946  & \textbf{0.6845 } & \textbf{0.5950 } & 0.6826  & 0.5915  & 0.6831  & 0.5919  \\
          & GraphCodeBert~\cite{guo2020graphcodebert} & \textbf{0.7303 } & \textbf{0.6477 } & 0.7291  & 0.6463  & 0.7195  & 0.6341  & 0.7203  & 0.6384  \\
          & DistilBert~\cite{sanh2019distilbert} & 0.5811  & 0.4822  & 0.5878  & 0.4928  & 0.5887  & 0.4963  & \textbf{0.5907 } & \textbf{0.4958 } \\
          \cmidrule{2-10}
          & UniXcoder~\cite{guo2022unixcoder} & 0.6698  & 0.5792  & 0.6652  & 0.5753  & \textbf{0.6749 } & \textbf{0.5875 } & 0.6704  & 0.5814  \\
          & SentenceBert~\cite{reimers2019sentencebert} & 0.6937  & 0.6055  & 0.6903  & 0.5989  & 0.6940  & 0.6025  & \textbf{0.6949 } & \textbf{0.6042 } \\
          & CoCoSoDa~\cite{shi2023cocosoda} & 0.6888  & 0.5954  & \textbf{0.6896 } & \textbf{0.5994 } & 0.6828  & 0.5862  & 0.6886  & 0.5954  \\
    \bottomrule
    \end{tabular}%
    }
  \label{tab:code_style}%
\end{table*}%

\subsubsection{\textbf{RQ3: How does the number of code styles generated based on logic-invariant strategy impact PseudoBridge's ability to align NL semantics and PL logic?}}

This experiment investigates the impact of code style diversity on the retrieval performance of PLMs. 
The core objective is to verify whether increasing the richness of code style variants, while maintaining logical equivalence, effectively enhances the model's capability to bridge the gap between NL semantics and PL logic.

\textbf{Setup.}
To assess the impact of code style diversity on PLM-based approaches' retrieval performance, we test how PseudoBridge performs with different numbers of code style variants ($k = 1,2,3,4$).
This experiment focuses on understanding whether increasing stylistic diversity in logically equivalent code enhances the model’s ability to bridge the semantic and logic gap between NL and PL.
To ensure a comprehensive evaluation, we select three representative baseline models from both the high/low-performing baselines, and evaluate them on the Python and Ruby datasets.

\textbf{Results and Analysis.}
Table~\ref{tab:code_style} shows that the number of style variants required for optimal retrieval varies across architectures and PLs. We derive the following observations:

\begin{itemize}
    \item \textbf{Differentiated improvement based on foundation model capabilities:} On the data-rich Python corpus, the marginal performance gains derived from style augmentation tend to diminish as the foundation model's intrinsic capabilities increase. For models with weaker representation capabilities, increasing style variants effectively compensates for deficiencies in their semantic representation space. For instance, when the number of style variants increases from 1 to 4, the MRR of DistilBERT steadily rises from 0.7772 to 0.7848. This indicates that multi-style training introduces diverse syntactic expressions. It helps bridge the alignment gap between NL semantics and PL logic. 
    In contrast, models with stronger semantic capture capabilities show clear diminishing returns. For example, SentenceBERT and CoCoSoDa reach peak performance with only one style variant. Increasing the number of variants further does not yield significant improvements. We even observe a slight decline for SentenceBERT. When $k=4$, the MRR drops to 0.8507. This suggests that such models already effectively capture the core logical invariants in the code. Excessive style augmentation may introduce redundant noise.
    \item \textbf{Distributional instability induced by data scarcity:} For the low-resource language Ruby, the impact of style augmentation does not follow a linear pattern. Instead, it exhibits significant distributional fluctuation. For instance, GraphCodeBERT reaches its peak MRR of 0.7303 at $k=1$. CoCoSoDa performs best at $k=2$ with a score of 0.6896. SentenceBERT requires an increase to $k=4$ to achieve an optimal result of 0.6949. This instability stems from the representation sparsity inherent in low-resource languages. The scarcity of training corpora makes it difficult for models to construct a continuous semantic space. Consequently, models tend to overfit surface syntactic patterns rather than capturing deep logical invariants. This tendency significantly weakens the model's robustness to code style transformations and triggers performance oscillations.
\end{itemize}

\begin{tcolorbox}[
    colback=gray!5!white,
    colframe=gray!95!black,
    fonttitle=\bfseries,
    arc=2.8pt,
    boxrule=0.9pt,
    enhanced,
    boxsep=1.5pt,              
    left=1.5pt, right=1.5pt,     
    top=1.5pt, bottom=1.5pt,     
    before upper={\setstretch{0.9}}
]
\textbf{Conclusion 3}: PseudoBridge enhances the alignment between natural language semantics and code logic by introducing diverse style variants. The extent of these gains depends on training corpus size and base model capability. In high-resource languages like Python, the benefit correlates negatively with model capability, as stronger models saturate with fewer variants. Conversely, in low-resource languages like Ruby, sparse representations compromise robustness, resulting in unstable correlations between variant quantity and retrieval performance.
\end{tcolorbox}

\begin{figure*}[t]
    \centering
    \includegraphics[width=0.9\linewidth]{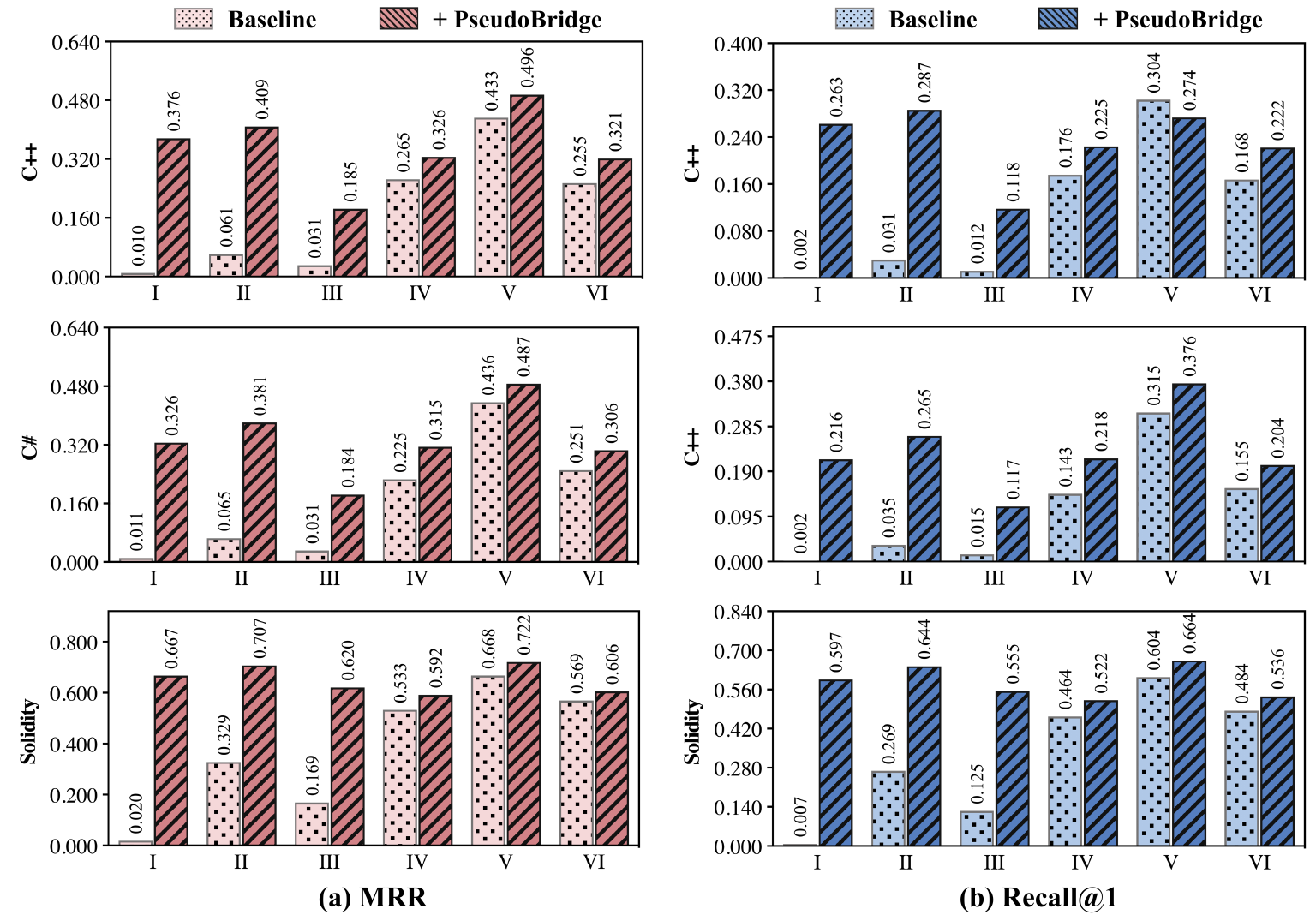}
    \caption{Zero-shot learning capability comparison across domains and tasks for baseline models with/without PseudoBridge fine-tuning \textbf{(RQ4)}. \textit{Note}: I: CodeBert; II: GraphCodeBert; III: DistilBert; IV: UniXcoder; V: SentenceBert; VI: CoCoSoDa.}
    \vspace{-0.15cm}
    \label{fig:cross_domain}
\end{figure*}

\subsubsection{\textbf{RQ4: How does PseudoBridge facilitate zero-shot knowledge transfer across domains and programming languages?}}
To systematically evaluate the zero-shot knowledge transfer ability of PseudoBridge in unseen programming languages and application domains, we conduct a cross-domain evaluation experiment. 
This experiment aims to discuss the universality of the alignment mechanism between NLs semantics and PLs logic. We examine whether the model effectively generalizes to different real scenarios without fine-tuning for the target language.

\textbf{Setup.}
We select three representative PLs to conduct zero-shot evaluations before and after fine-tuning. The models are not exposed to data related to the target languages during the training phase. These datasets cover diverse programming paradigms and application scenarios:

\begin{itemize}
    \item \textbf{General-Purpose Languages:} We select \textbf{C++} and \textbf{C\#} as representatives. The former embodies the complexity of low-level resource management and high-performance computing. The latter represents the strongly-typed object-oriented paradigm typical of enterprise systems. These experiments aim to evaluate the model's generalization capability across different programming paradigms within mainstream software engineering scenarios.
    \item \textbf{Domain-Specific Language:} We choose \textbf{Solidity} to represent this category. As the core language for smart contract development, it involves unique logic regarding blockchain transactions and security constraints. These experiments aim to verify whether the model maintains robust adaptability in vertical domain scenarios characterized by data scarcity and heterogeneous logic.
\end{itemize}

\textbf{Results and Analysis.} 
As shown in Figure~\ref{fig:cross_domain}, the experimental results indicate that models fine-tuned with PseudoBridge achieve significant improvements in code retrieval accuracy. We derive the following key findings:

\begin{itemize}
    \item \textbf{Superior cross-language generalization:} As shown in Figure~\ref{fig:cross_domain}, PseudoBridge yields significant performance gains across six PLMs. This occurs even when models are not exposed to data from target languages during training. Specifically, CodeBERT shows low initial performance on C++ and C\# datasets, with scores of 0.0022 and 0.002, respectively. After fine-tuning with PseudoBridge, its performance surges to 0.2625 and 0.2156. Even UniXcoder, which exhibits better baseline performance, achieves absolute improvements of 0.0489 on C++ and 0.0748 on C\#. This indicates that PseudoBridge captures deep, transferable semantic representations of code logic rather than surface-level syntactic features.
    \item \textbf{Broad model and scenario adaptability:} Experimental results indicate that this method significantly enhances PLMs with varying capabilities, particularly those with weaker baselines. In tasks involving general-purpose languages such as C++ and C\#, most models achieve substantial gains. For instance, the performance of DistilBERT on C++ increases nearly 10-fold, rising from 0.0122 to 0.1179. Similarly, the MRR of GraphCodeBERT on C\# increases by 0.23 to reach 0.2651. In domain-specific language tasks like Solidity, the models demonstrate strong adaptability. CodeBERT surges from 0.007 to 0.597, while SentenceBERT achieves steady growth from 0.604 to 0.664 despite its strong baseline. This universal improvement across diverse architectures including BERT-based models and UniXcoder effectively bridges the semantic gap, confirming the robustness of the model in heterogeneous scenarios.
\end{itemize}

\begin{tcolorbox}[
    colback=gray!5!white,
    colframe=gray!95!black,
    fonttitle=\bfseries,
    arc=2.8pt,
    boxrule=0.9pt,
    enhanced,
    boxsep=1.5pt,              
    left=1.5pt, right=1.5pt,     
    top=1.5pt, bottom=1.5pt,     
    before upper={\setstretch{0.9}}
]
\textbf{Conclusion 4}: PseudoBridge exhibits strong zero-shot transfer capabilities across both programming languages and application domains. This strong generalization performance suggests that the core mechanism of aligning NL semantics with PL logic plays a pivotal role in enabling effective cross-domain knowledge transfer. Consequently, it provides a scalable solution for code retrieval in heterogeneous software engineering environments.
\end{tcolorbox}

\begin{figure*}[t]
    \centering
    \includegraphics[width=0.88\linewidth]{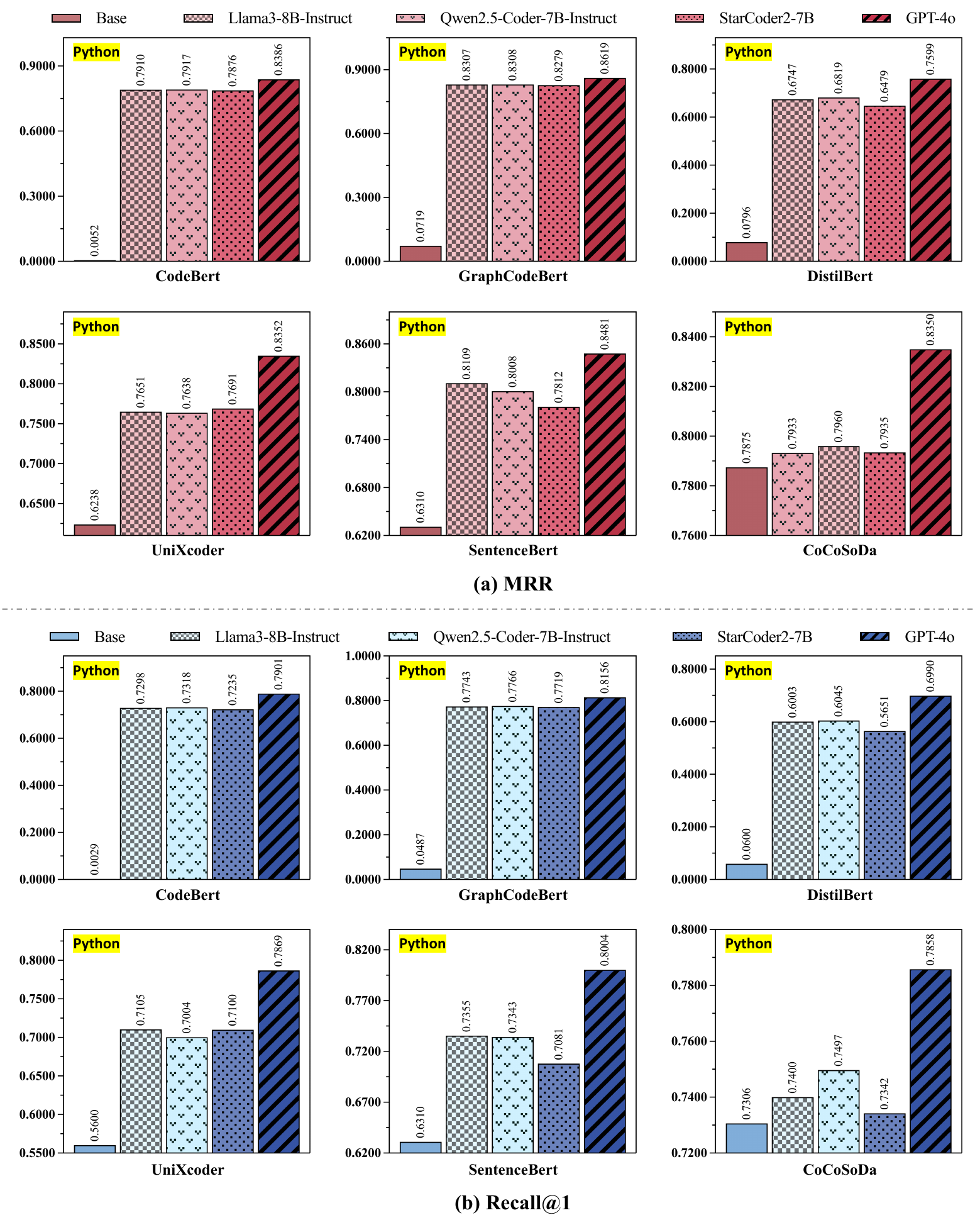}
    \caption{Comparison of performance improvements for six baseline models on the Python dataset \textbf{(RQ5)}. Pseudo-code and multi-style code are generated utilizing \textit{StarCoder2-7B}, \textit{Llama3-8B-Instruct}, \textit{Qwen2.5-Coder-7B-Instruct}, and \textit{GPT-4o}.}
    \vspace{-0.15cm}
    \label{new_figs:open_model_python}
\end{figure*}

\subsubsection{\textbf{RQ5: How robust is PseudoBridge to the capabilities of the data synthesis model?}}

To determine whether the performance gains stem from the PseudoBridge framework itself or primarily from the high-quality data generated by GPT-4o, we employ open-source LLM as synthesis engines for comparative analysis. To simulate real-world scenarios with limited resources, we fine-tune the baseline models using only 20\% of the synthetic data.

\textbf{Setup.}
To systematically verify the robustness of our framework across varying generation capabilities, we select three open-source LLMs with parameter sizes ranging from 7B to 8B as synthesis engines. These models cover diverse architectural paradigms and capability dimensions, as detailed below.

\begin{itemize}
    \item \textbf{Code Foundation Models:} We use \textbf{\textit{StarCoder2-7B}}~\cite{lozhkov2024starcoder} to represent code foundation models pre-trained on the Stack v2 code corpus in an unsupervised manner. This model serves to demonstrate the robustness of our framework when using a code foundation model without extensive instruction fine-tuning.
    \item \textbf{General-Purpose LLMs:} We select \textbf{\textit{Llama-3-8B-Instruct}}~\cite{grattafiori2024llama}, a general-purpose LLM offering robust language understanding despite lacking code-specific optimization. We include this model to investigate whether PseudoBridge leverages general logical reasoning for effective code intent alignment in non-specialized models.
    \item \textbf{Instruction-Tuned Code Models:} We employ \textbf{\textit{Qwen2.5-Coder-7B-Instruct}}~\cite{hui2024qwen2coder} as a representative instruction-tuned model, noted for its superior code generation and instruction-following capabilities. This allows us to examine how the synthesis quality gap between open-source and closed-source models affects retrieval.
\end{itemize}

\begin{figure*}[t]
    \centering
    \includegraphics[width=0.88\linewidth]{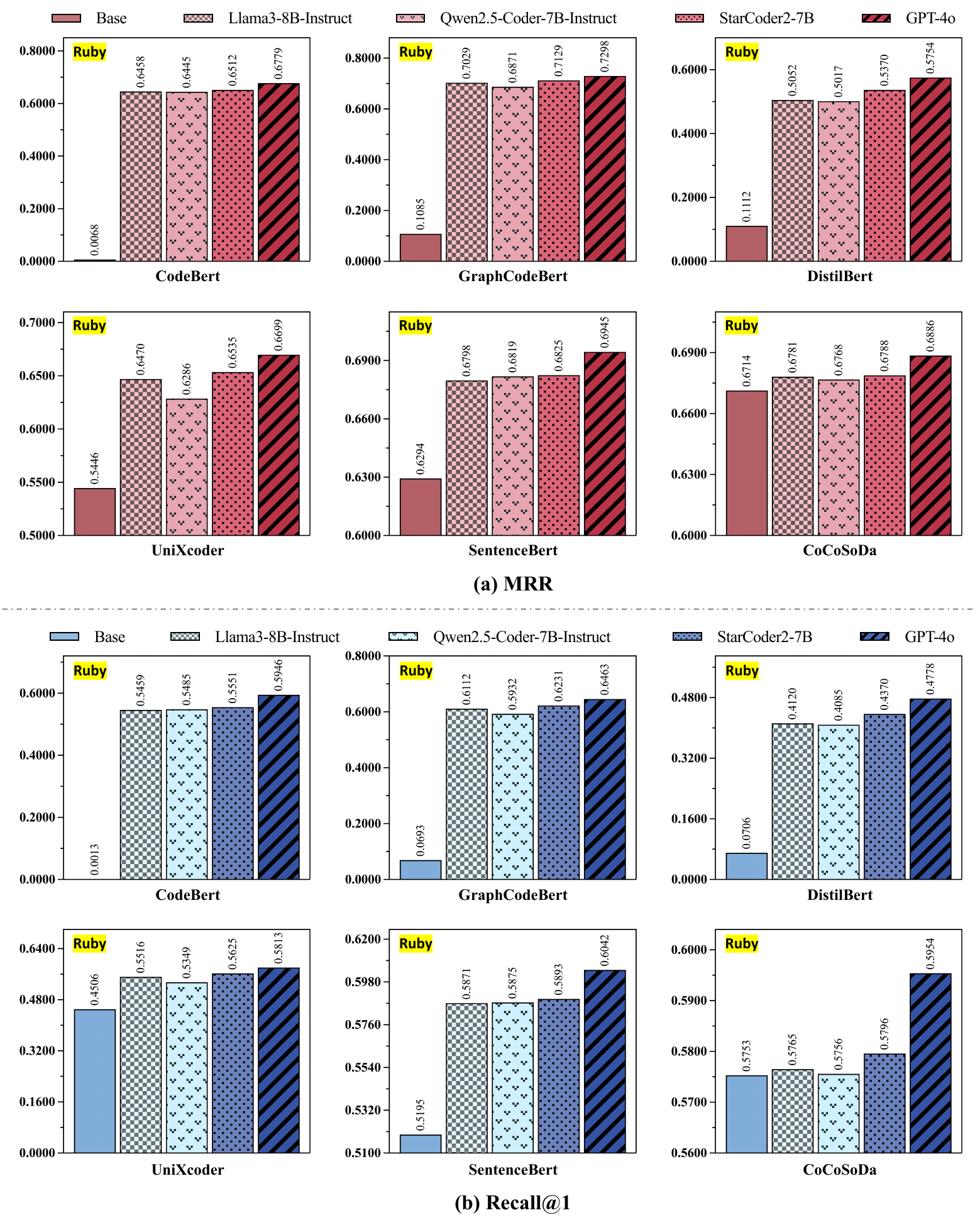}
    \caption{Comparison of performance improvements for six baseline models on the Ruby dataset \textbf{(RQ5)}. Pseudo-code and multi-style code are generated utilizing \textit{StarCoder2-7B}, \textit{Llama3-8B-Instruct}, \textit{Qwen2.5-Coder-7B-Instruct}, and \textit{GPT-4o}.}
    \vspace{-0.15cm}
    \label{new_figs:open_model_ruby}
\end{figure*}

\textbf{Results and Analysis.}
Figures~\ref{new_figs:open_model_python} and \ref{new_figs:open_model_ruby} illustrate the retrieval performance comparisons of different generators across various baseline models. We observe the following key findings from these results.

\begin{itemize}
    \item \textbf{Significant performance improvements:} Taking the Python dataset as an example, as shown in Figure~\ref{new_figs:open_model_python}, all three open-source models demonstrate substantial gains across all six baseline models. 
    Even when using the basic \textbf{\textit{StarCoder2-7B}} to generate data, the MRR of CodeBERT surges from an initial 0.0052 to 0.7876, and the Recall@1 of GraphCodeBERT improves from 0.0487 to 0.7719. 
    A similar trend appears in the Figure~\ref{new_figs:open_model_ruby} Ruby dataset. 
    These results strongly prove that the performance improvement primarily stems from the core mechanism of using ``pseudo-code as an intermediate bridge''. 
    This mechanism effectively helps the model capture the deep logical intent of the code, and its effectiveness does not rely entirely on advanced LLMs.
    \item \textbf{Dominance of framework effectiveness:} 
    Comparing different open-source generators, \textbf{\textit{Qwen2.5-Coder-7B-Instruct}} and \textbf{\textit{Llama-3-8B-Instruct}} perform better than \textbf{\textit{StarCoder2-7B}} in most cases. 
    This indicates that as the coding capability of the generator improves, the benefits of PseudoBridge can grow further. 
    However, the incremental gain from upgrading from a 7B open-source model to \textbf{\textit{GPT-4o}} is far smaller than the massive performance leap achieved by moving from ``w/o PseudoBridge'' to ``+ PseudoBridge''. This demonstrates that the PseudoBridge framework design itself is the primary source of effectiveness.
\end{itemize}

\begin{tcolorbox}[
    colback=gray!5!white,
    colframe=gray!95!black,
    fonttitle=\bfseries,
    arc=2.8pt,
    boxrule=0.9pt,
    enhanced,
    boxsep=1.5pt,              
    left=1.5pt, right=1.5pt,     
    top=1.5pt, bottom=1.5pt,     
    before upper={\setstretch{0.9}}
]
\textbf{Conclusion 5}: PseudoBridge exhibits strong robustness regarding the capabilities of the generator. Even when using low-cost, small-scale open-source models like \textit{Qwen2.5-Coder-7B-Instruct} with only 20\% of the training data, PseudoBridge still helps retrieval models achieve significant performance gains. This finding confirms that the core strength of the framework lies in its alignment mechanism rather than a specific data source. It also demonstrates the broad potential of this method for private deployment and resource-constrained scenarios.
\end{tcolorbox}

\subsubsection{\textbf{RQ6: How does PseudoBridge compare with modern LLMs and specialized embedding baselines?}}
To assess the practical utility of PseudoBridge, we compare it with representative large-scale models. Unlike traditional PLMs, modern LLMs and specialized embedding services feature massive parameter scales and vast pre-training datasets. This experiment investigates whether PseudoBridge enables traditional PLMs to achieve competitive performance when facing these resource-intensive baselines.

\textbf{Setup.}
We conduct evaluations on the complete test sets for Python and Ruby. To compare performance across different technical approaches, we include the following representative models as baselines:

\begin{itemize}
    \item \textbf{Traditional Pre-trained Language Models:} We adopt GraphCodeBERT and use the model fine-tuned with the PseudoBridge method as a basic baseline.
    \item \textbf{Specialized Embedding Models:} We choose \textit{\textbf{Qwen3-Embedding-8B}}~\cite{qwen3embedding} as the open-source representative and \textit{\textbf{Text-Embedding-3-Small}} as the closed-source representative. The former is designed for retrieval tasks. The latter is a widely used service from OpenAI that serves as the industrial benchmark for dense retrieval.
    \item \textbf{Generative Code LLMs:} We include \textit{\textbf{Qwen2.5-Coder-7B}} and its instruction-tuned variant, \textit{\textbf{Qwen2.5-Coder-7B-Instruct}}~\cite{hui2024qwen2coder}. While these models excel in code reasoning and generation, they are not inherently optimized for embedding or retrieval tasks. We select them to evaluate whether advanced generative proficiency directly translates into effective semantic retrieval capabilities.
\end{itemize}

\begin{figure}[htbp]
    \centering
    \includegraphics[width=0.88\linewidth]{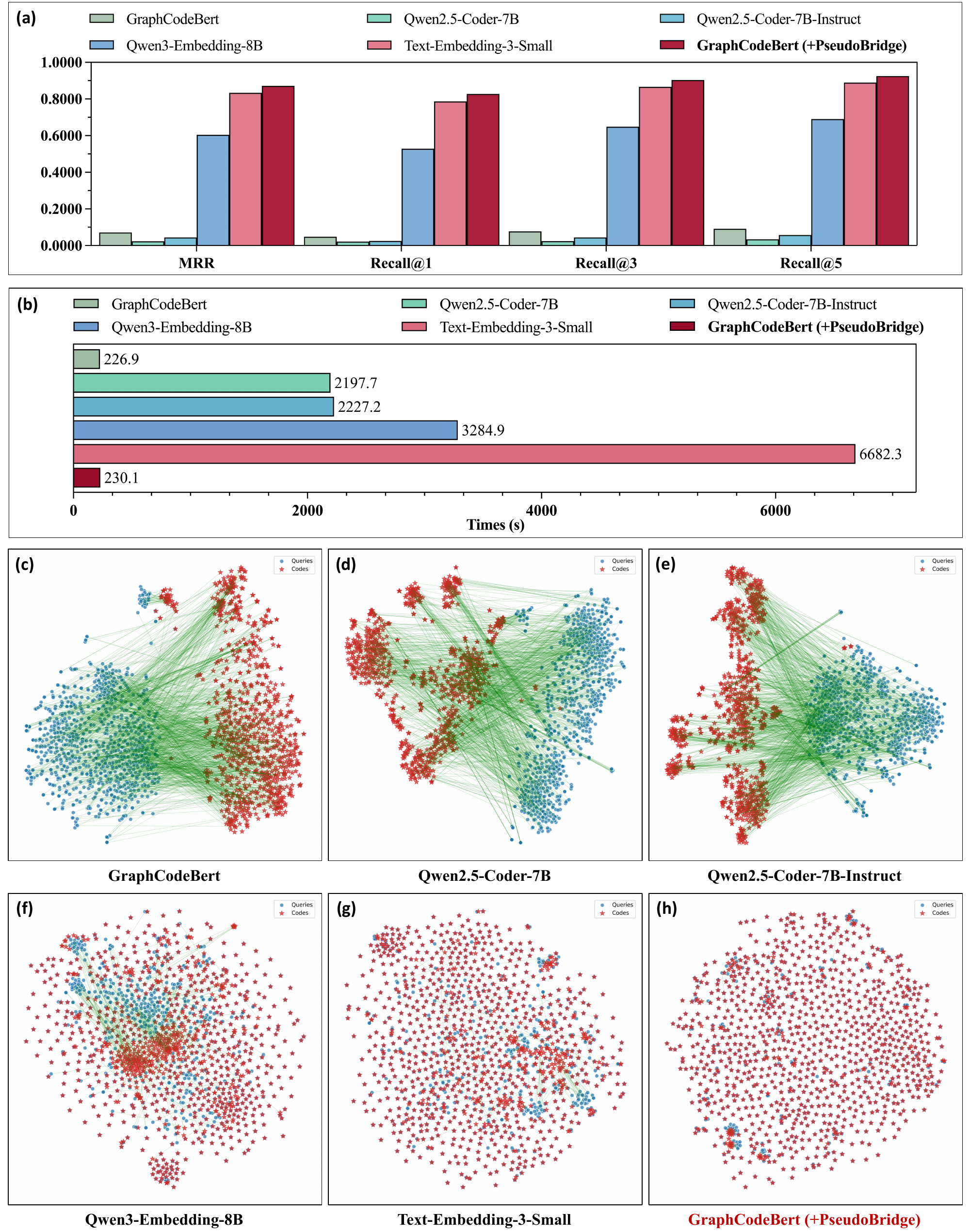}
    \caption{Performance and efficiency comparison with modern LLM-based embedding baselines on the Python dataset \textbf{(RQ6)}. (a) Comparison of retrieval metrics (MRR, Recall@1, 3, 5). (b) Comparison of inference time costs. (c)--(h) t-SNE visualizations of the embedding spaces for GraphCodeBERT, Generative Code LLMs (\textit{Qwen2.5-Coder-7b}, \textit{Qwen2.5-Coder-7b-Instruct}), Specialized Embedding Models (\textit{Qwen3-Embedding-8B}, \textit{Text-Embedding-3-Small}), and GraphCodeBERT (+PseudoBridge), respectively. In the t-SNE plots, green lines connect NL and PL pairs; longer lines indicate larger semantic distances and lower retrieval accuracy.}
    \label{new_figs:python_llms_embedding}
\end{figure}

\begin{figure}[htbp]
    \centering
    \includegraphics[width=0.88\linewidth]{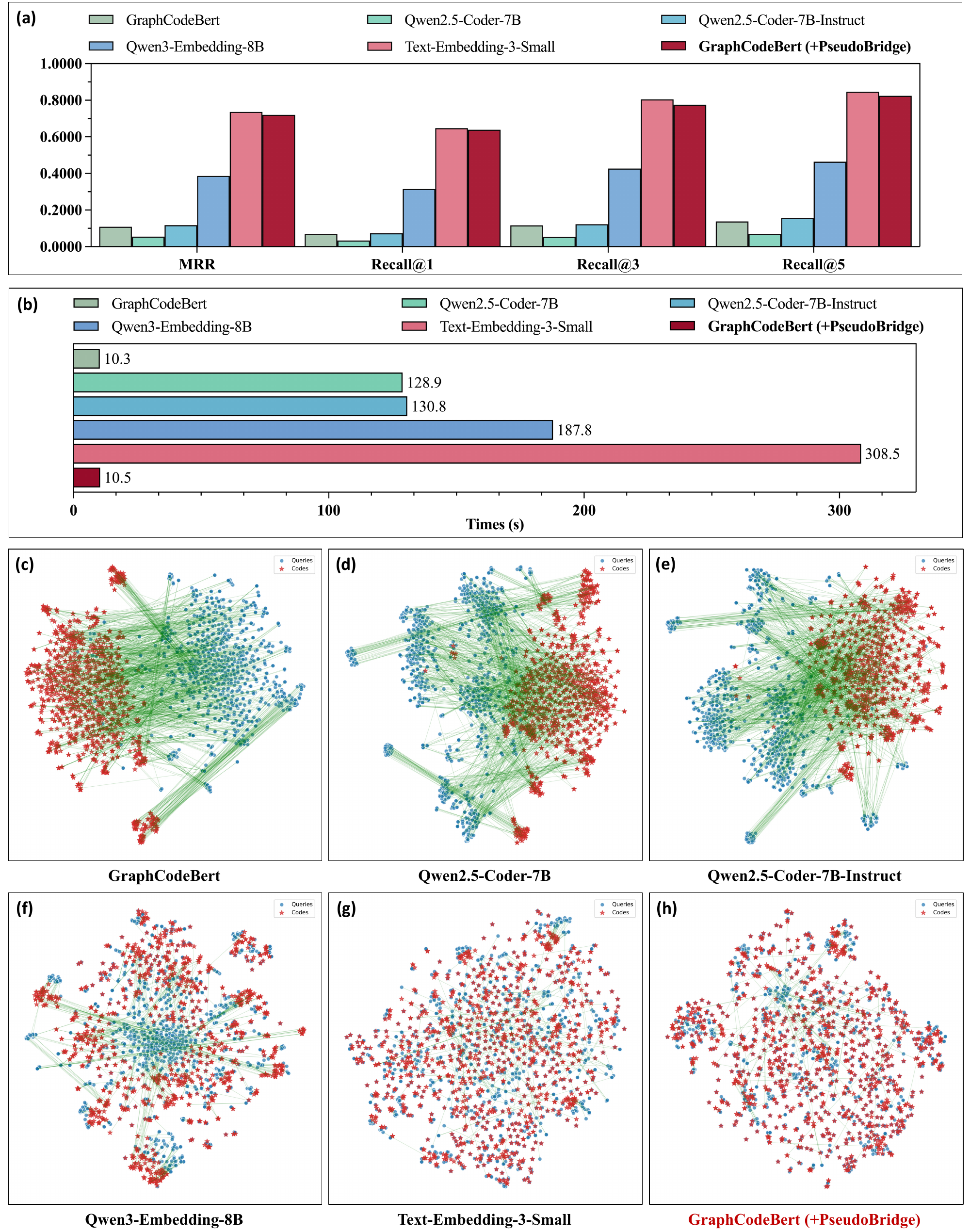}
    \caption{Performance and efficiency comparison with modern LLM-based embedding baselines on the Ruby dataset \textbf{(RQ6)}. (a) Comparison of retrieval metrics (MRR, Recall@1, 3, 5). (b) Comparison of inference time costs. (c)--(h) t-SNE visualizations of the embedding spaces for GraphCodeBERT, Generative Code LLMs (\textit{Qwen2.5-Coder-7b}, \textit{Qwen2.5-Coder-7b-Instruct}), Specialized Embedding Models (\textit{Qwen3-Embedding-8B}, \textit{Text-Embedding-3-Small}), and GraphCodeBERT (+PseudoBridge), respectively. In the t-SNE plots, green lines connect NL and PL pairs; longer lines indicate larger semantic distances and lower retrieval accuracy.}
    \label{new_figs:ruby_llms_embedding}
\end{figure}

\textbf{Results and Analysis.}
Figure~\ref{new_figs:python_llms_embedding} and Figure~\ref{new_figs:ruby_llms_embedding} present the experimental results. Based on the comparative analysis, we draw the following key conclusions:

\begin{itemize}
    \item \textbf{Decoupling of generation and retrieval capabilities:} Experiments show that while generative models like \textbf{\textit{Qwen2.5-Coder}} excel in code synthesis, their zero-shot retrieval performance is very limited (MRR < 0.05). This gap reveals that strong logic reasoning does not automatically yield high-quality retrieval embeddings. As qualitatively evidenced by the t-SNE visualizations in Figure~\ref{new_figs:python_llms_embedding}(d)-(e), the embedding spaces of generative LLMs exhibit scattered distributions and elongated semantic distances between NL-PL pairs. PseudoBridge addresses this by converting large-model knowledge into retrieval features through explicit logic and semantic alignment.
    \item \textbf{Optimization of performance and resource efficiency:}Traditional PLMs fine-tuned with PseudoBridge achieve an MRR of 0.8721 on Python, surpassing \textbf{\textit{Qwen3-Embedding-8B}} (MRR 0.8339) and rivaling \textbf{\textit{Text-Embedding-3-Small}}. While the reduction in inference time (from 2,197.7s to 230.1s) is attributed to the lightweight 125M architecture, the significance of PseudoBridge lies in enabling such a compact model to bridge the performance gap. As shown in Figure~\ref{new_figs:python_llms_embedding}(h), PseudoBridge significantly tightens the alignment compared to the original GraphCodeBERT (Figure~\ref{new_figs:python_llms_embedding}(c)), proving that explicit alignment is essential for achieving high precision in resource-constrained scenarios.
\end{itemize}

\begin{tcolorbox}[
    colback=gray!5!white,
    colframe=gray!95!black,
    fonttitle=\bfseries,
    arc=2.8pt,
    boxrule=0.9pt,
    enhanced,
    boxsep=1.5pt,              
    left=1.5pt, right=1.5pt,     
    top=1.5pt, bottom=1.5pt,     
    before upper={\setstretch{0.9}}
]
\textbf{Conclusion 6}: PseudoBridge shows that explicit logic-semantic alignment enables traditional PLMs to match the accuracy of 8B-scale specialized models and industrial APIs. Its combination of competitive performance and low inference latency offers a resource-efficient and viable solution for large-scale industrial code retrieval.
\end{tcolorbox}

\subsection{Ablation Studies}

\subsubsection{Ablation Study on PseudoBridge Components} 
We conduct comprehensive ablation studies on six baseline models using Python and Ruby datasets to analyze the specific contributions of individual components within the PseudoBridge framework to code retrieval performance. 
This study validates the effectiveness of two core designs, which are pseudo-code as an intermediate semantic bridge and multi-style code augmentation. 
The experiment addresses two primary research questions. 
First, we investigate whether pseudo-code plays an indispensable role in bridging natural language and programming language. 
Second, we examine whether the logically invariant code style augmentation strategy effectively improves the generalization robustness of the model.

\textbf{Setup.}
We design three variants to evaluate the impact of specific components by removing them. The detailed information as follows:

\begin{itemize}
    \item \textit{\textbf{w/o Stage 1}}: In this configuration, we remove the <NL, Pseudo-code> alignment training. We proceed directly to the second stage of training using the generated pseudo-code and multi-style code. This setup evaluates whether the model learns effective representations using only code variants when pseudo-code does not serve as a direct semantic continuation of natural language.
    \item \textit{\textbf{w/o Stage 2}}: In this setting, we fine-tune the model exclusively on <NL, Pseudo-code> pairs. The model does not access real codes. This configuration verifies whether retrieval performance suffers when aligning queries solely with pseudo-code, specifically examining the impact of lacking grounding in executable code.
    \item \textit{\textbf{w/o Stage Code Style}}: In this setting, we retain the two-stage training process but utilize only <Pseudo-code, Original Code> during the second stage. We exclude the multi-style code variants generated by large language models. This configuration quantifies the contribution of code style diversity to mitigating model overfitting on surface syntactic features.
\end{itemize}

\begin{figure}[htbp]
    \centering
    \includegraphics[width=0.9\linewidth]{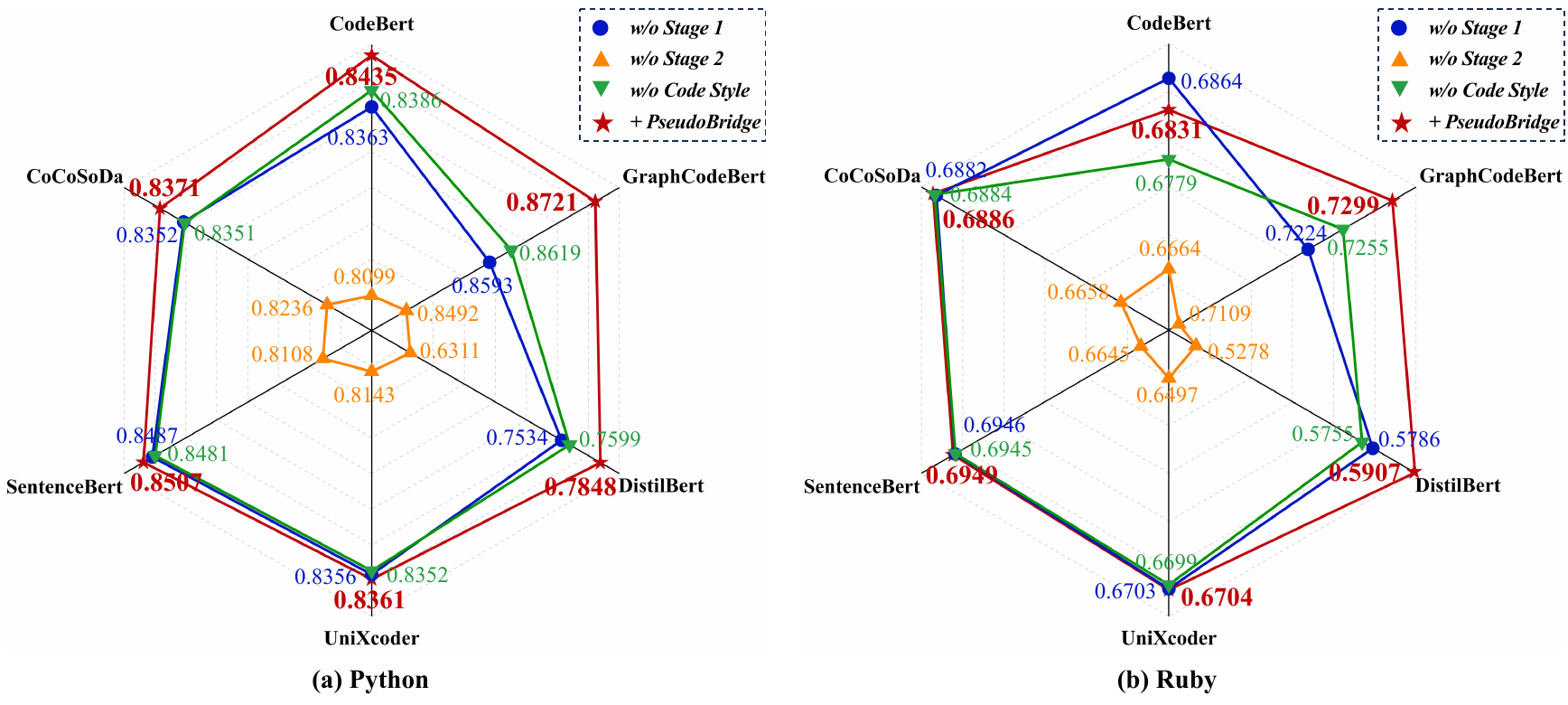}
    \caption{Ablation study on the component analysis based on Python and Ruby. \textit{Note}: (1) \textit{\textbf{w/o Stage 1}}: removes <NL,Pseudo-code> pairs and directly trains <Pseudo-code, Multi-style code variants>; (2) \textit{\textbf{w/o Stage 2}} only uses <NL,Pseudo-code> pairs training, without original code and multi-style code variants; (3) \textit{\textbf{w/o Code Style}} trains on <NL,Pseudo-code>, then on <Pseudo-code, Original code>, without style-enhanced variants; (4) \textit{\textbf{+ PseudoBridge}} employs training with pseudo-code to natural language pairs, then on pseudo-code to style-enhanced code variants.}
    \vspace{-0.15cm}
    \label{fig:ablation_studies}
\end{figure}

\textbf{Results and Analysis.}
Figure~\ref{fig:ablation_studies} presents the experimental results. We compare the performance impact of removing different components and analyze the sensitivity differences across various model architectures. We derive some interesting results from these observations.

\begin{itemize}
    \item \textbf{Important Aligning Pseudo-Code with Programming Languages.} The configuration without \textit{\textbf{Stage 2}} consistently performs the worst across all settings. For instance, DistilBERT's MRR on the Python dataset drops from 0.7848 to 0.6311 ($\downarrow 20\%$). This indicates that pseudo-code acts as a semi-structured intermediate representation. While mapping natural language to pseudo-code captures algorithmic intent, it loses specific implementation details. \textit{\textbf{Stage 2}} grounds this abstract logic into concrete implementations; its absence causes a severe semantic disconnection, leading to ineffective retrieval.
    \item \textbf{Differential Effectiveness on Low/High-Performance Models.} 
    The performance of the text-only model DistilBERT on the Python dataset drops significantly from 0.7848 to 0.7534 when without \textit{\textbf{Stage 1}}. This outcome shows that text-only models rely heavily on pseudo-code to construct logical backbones because they cannot parse code structures effectively. Pseudo-code acts as a tool to provide structured knowledge. In contrast, high-performance models like UniXcoder and CoCoSoDa already learn structural information during pre-training through tasks like abstract syntax trees or data flow. These models depend less on external logical bridges, so they get less extra benefit. The main value of PseudoBridge is that it fills the gap in structural understanding for base models while providing better structural constraints for high-performance models.
\end{itemize}

\begin{tcolorbox}[
    colback=gray!5!white,
    colframe=gray!95!black,
    fonttitle=\bfseries,
    arc=2.8pt,
    boxrule=0.9pt,
    enhanced,
    boxsep=1.5pt,              
    left=1.5pt, right=1.5pt,     
    top=1.5pt, bottom=1.5pt,     
    before upper={\setstretch{0.9}}
]
\textbf{Finding 1}: PseudoBridge introduces pseudo-code and code style enhancements based on logical invariant strategy, both of which are crucial for improving the code retrieval performance of baseline models. This two-stage design effectively bridges logic and semantic alignment between NL and PL, enabling the model to capture functional intent through pseudo-code alignment before training to enhance its generalization capabilities.
\end{tcolorbox}

\subsubsection{Representation Impact: Pseudo-code vs. SCoT vs. ASTs}

\begin{figure}[htbp]
    \centering
    \includegraphics[width=0.88\linewidth]{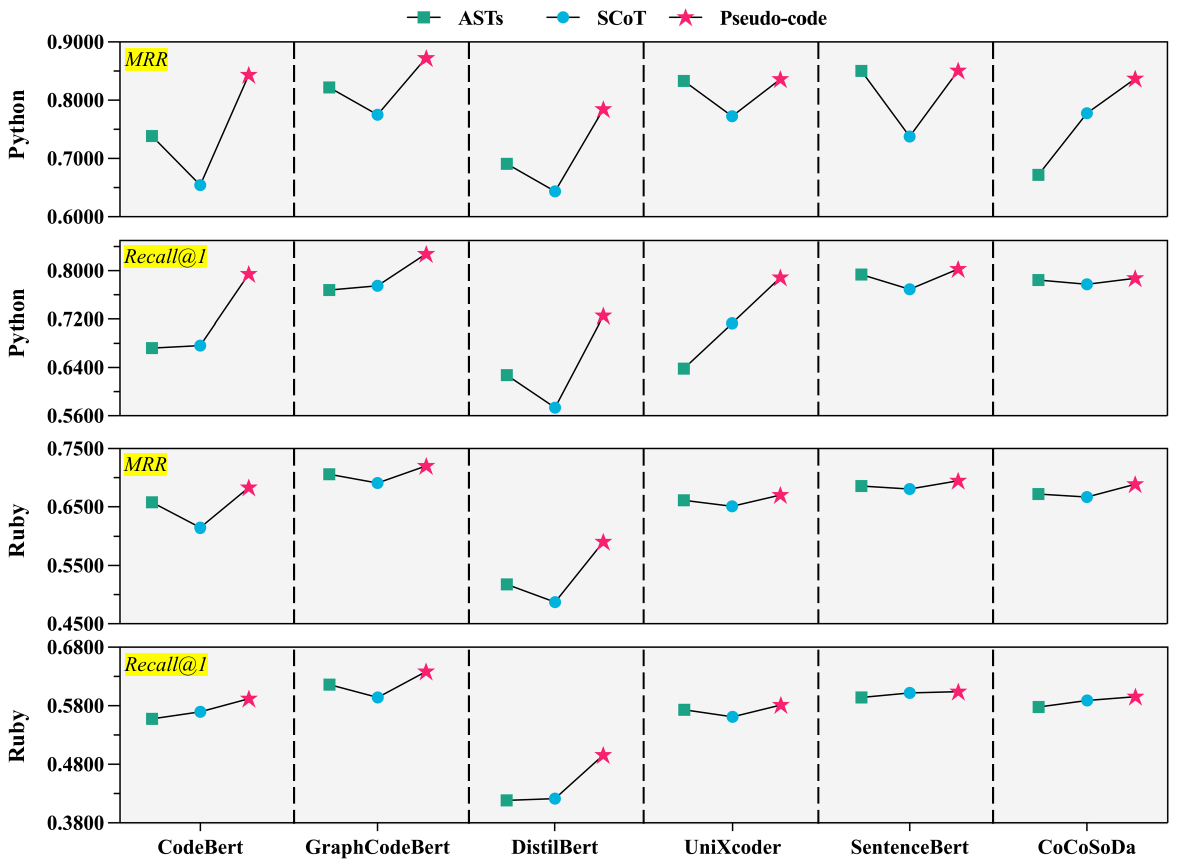}
    \caption{Ablation study on Pseudo-code vs. ASTs intermediate representations for code retrieval.}
    \label{new_figs:scot_ast_pseudo_compare}
\end{figure}

To examine the impact of different Intermediate Representations (IRs) on code retrieval performance, we design a comparative ablation study. 
Specifically, we replace the pseudo-code in PseudoBridge with two representative IRs: (1) Abstract Syntax Trees (ASTs) , which represents the underlying syntactic structure of the code; and (2) Structured Chain-of-Thought (SCoT)~\cite{li2025scot}, which describes the reasoning process for code generation tasks. 
We conduct experiments on the Ruby and Python datasets and perform a comprehensive evaluation using six backbone models. Figure~\ref{new_figs:scot_ast_pseudo_compare} shows that pseudo-code consistently achieves the best performance across all models and datasets. Detailed analysis as follows:

\begin{itemize}
    \item \textbf{Superiority of Pseudo-code: Logical and Semantic Alignment.} Experimental results demonstrate that while introducing ASTs or SCoT yields performance gains over baselines by providing structural insights, pseudo-code consistently outperforms them across all metrics. This confirms that pseudo-code functions as an optimal semantic anchor bridging NL intent and PL implementation. Rather than merely increasing data diversity, it facilitates deep cross-modal alignment by abstracting away syntactic noise and procedural details.
    \item \textbf{Limitations of ASTs: Syntactic Noise and Semantic Gap.} Although ASTs provide rich hierarchical structures, their retrieval performance generally falls short of pseudo-code. ASTs contain language-specific syntactic details, such as node types and nesting rules, that are irrelevant to functional intent and act as syntactic noise. This rigid structure compels models to overfit to implementation forms rather than focusing on high-level semantics. Consequently, this widens the semantic gap between natural language queries and code, thereby limiting model generalization in heterogeneous spaces.
    \item \textbf{Limitations of SCoT: Process Redundancy and Goal Misalignment.} SCoT facilitates code generation through explicit steps involving IO, sequence, branching, and loops. However, it exhibits suboptimal performance in retrieval tasks, with metrics occasionally falling below those of ASTs. Fundamentally, SCoT represents a process-oriented reasoning trace. Its verbose steps introduce significant procedural details and reasoning artifacts that often diverge from the core functional semantics of a user query. This shifts the focus from ``what the code does'' to ``how to generate it'', thereby introducing noise into the vector space and degrading retrieval precision.
\end{itemize}

\begin{tcolorbox}[
    colback=gray!5!white,
    colframe=gray!95!black,
    fonttitle=\bfseries,
    arc=2.8pt,
    boxrule=0.9pt,
    enhanced,
    boxsep=1.5pt,              
    left=1.5pt, right=1.5pt,     
    top=1.5pt, bottom=1.5pt,     
    before upper={\setstretch{0.9}}
]
\textbf{Finding 2}: 
Pseudo-code functions as a preferable intermediate representation for code retrieval tasks and outperforms structured or procedural alternatives. ASTs is susceptible to noise from language-specific syntax, while SCoT contains procedural redundancy irrelevant to the retrieval objective. In contrast, pseudo-code acts as a logical invariant. It strips away implementation details and preserves the core algorithmic skeleton to establish a robust semantic bridge between NL intent and PL logic.
\end{tcolorbox}

\subsubsection{Mechanism Analysis: Identifying the True Source of Performance Gains}
To identify the true source of performance gains in PseudoBridge, we construct two controlled baseline variants, namely Direct NL-PL and NL Paraphrase. By excluding pseudo-code while keeping the training volume constant, this design isolates our core variables. It allows us to determine whether the improvements stem from the explicit logical and semantic structures of pseudo-code, or simply from increased data scale and semantic diversity.

\textbf{Setup.} To completely exclude the impact of data scale, we strictly set the total volume of both baselines to 45,000 instances (19,000 from Stage 1, and 26,000 from Stage 2), aligning with the exact scale of PseudoBridge. The specific configurations are as follows:

\begin{itemize}
    \item \textit{\textbf{Direct NL-PL:}} We expand the NL-PL pairs from 19,000 to 45,000 by sampling from the filtered candidate set according to the criteria and language ratios in Section~\ref{section:dataset}. This configuration does not use any intermediate representation and skips the two-stage training framework. It examines whether increasing the data scale alone allows the model to reach a performance level similar to PseudoBridge without logical and semantic alignment.
    \item \textit{\textbf{NL Paraphrase:}} Following the paraphrasing concepts in the ReCode framework~\cite{wang2023recode}, this configuration employs \textit{\textbf{GPT-4o}} (\textit{version: gpt-4o-2024-08-06}) to semantically rewrite original queries, generating 19,000 unstructured NL paraphrases to serve as the IR. We strictly follow the two-stage training and data scale of PseudoBridge, only replacing the pseudo-code with these NL paraphrased texts. Stage 1 aligns original queries with NL paraphrases, and Stage 2 aligns NL paraphrases with code variants. This configuration explores whether the improvements are driven by the special structure of pseudo-code or simply by arbitrary training signals.
\end{itemize}

\begin{figure}[htbp]
    \centering
    \includegraphics[width=0.8\linewidth]{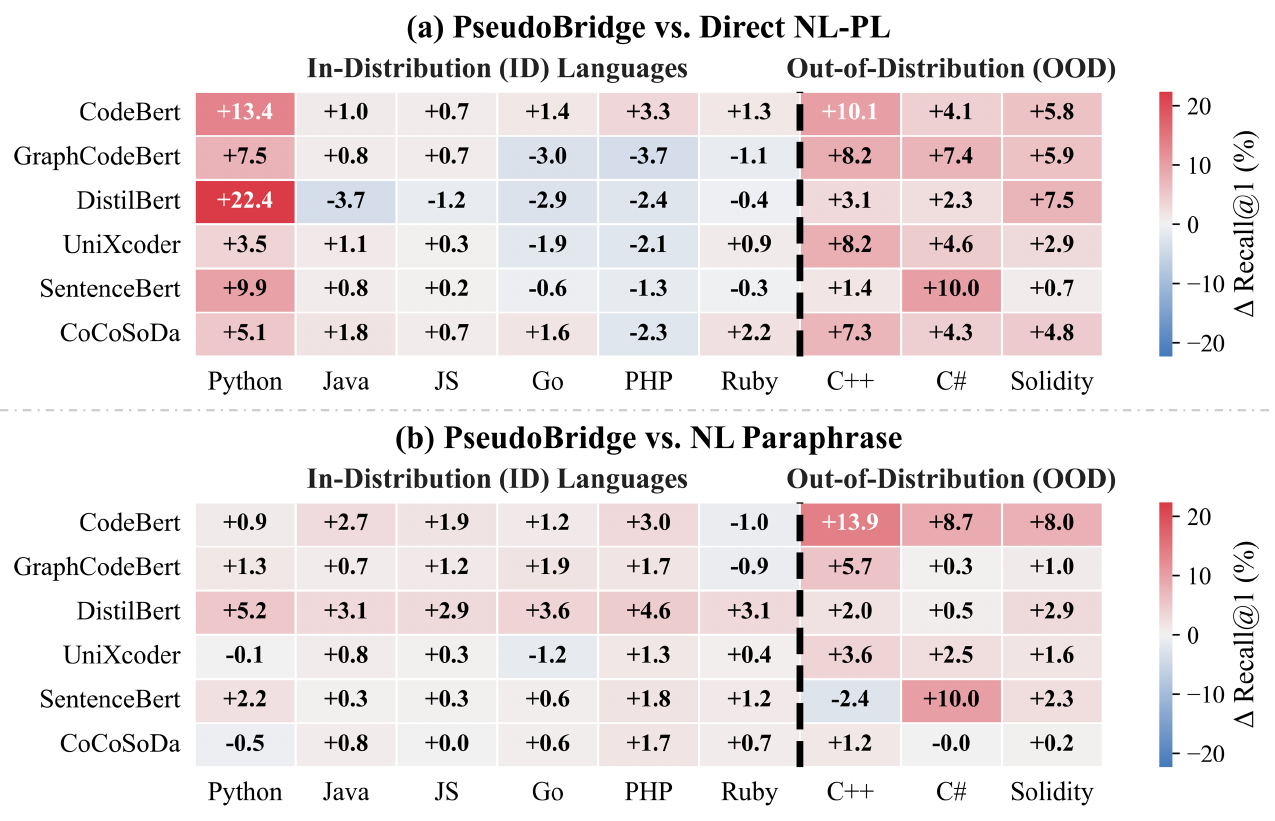}
    \caption{Heatmaps illustrating the absolute improvement ($\Delta$ Recall@1, \%) of PseudoBridge compared to two scale-matched baselines across 6 backbones and 9 PLs. (a) Comparison with \textit{Direct NL-PL} isolates the effect of structured intermediate representation versus pure data scaling. (b) Comparison with \textit{NL Paraphrase} isolates the importance of explicit logic abstraction versus unstructured semantic perturbation. The dashed vertical line separates In-Distribution (ID) languages from Out-of-Distribution (OOD) zero-shot languages. Red cells indicate positive gains by PseudoBridge.}
    \label{new_figs:nl_rewrite_direct}
  \vspace{-0.15cm}
\end{figure}

\textbf{Results and Analysis.} Figure~\ref{new_figs:nl_rewrite_direct} presents the experimental results. Across 54 configurations (9 PLs $\times$ 6 models), PseudoBridge achieves the best performance in 35 (64.8\%), and notably in 16 of 18 OOD configurations (88.9\%), whereas \textit{NL Paraphrase} wins only 5 (9.3\%). We analyze the results along the following dimensions:

\begin{itemize}
    \item \textbf{Data scaling cannot replace the explicit alignment from pseudo-code.} While \textbf{PseudoBridge} generally outperforms \textit{\textbf{Direct NL-PL}} on mainstream ID languages, its slight lag in Go and Ruby reflects a strategic trade-off in representation abstraction. By stripping syntactic noise, \textbf{PseudoBridge} prioritizes core logic, whereas \textit{\textbf{Direct NL-PL}} relies on language-specific ``shortcut mappings.'' These surface features offer minor ID gains but lack the robustness required for logical variations or cross-language transfer. OOD tests validate that baselines suffer from feature space shifts due to syntactic fitting, while \textbf{PseudoBridge}'s invariant logical anchors ensure robust generalization. For example, CodeBERT's $Recall@1$ on C++ improves from 0.1617 to 0.2625, proving that mapping NL to universal algorithmic structures yields generalization unattainable by simple syntactic stacking.
    \item \textbf{Gains depend on the specific structure of pseudo-code rather than arbitrary additional signals.} Experimental results show that \textit{\textbf{NL Paraphrase}} consistently underperforms \textbf{PseudoBridge} and often fails to exceed the baseline in ID scenarios, even causing degradation in Ruby and C\#. This indicates that semantic diversity alone, lacking logical constraints, leads to surface-pattern fitting rather than deep retrieval understanding. This gap is further highlighted in zero-shot OOD tests (C++, C\#, and Solidity), where \textit{\textbf{NL Paraphrase}} lacks the structural anchors necessary for generalization. Conversely, PseudoBridge demonstrates superior robustness by aligning NL with algorithmic logic and semantics, achieving a 10.1\% increase in Recall@1 for C++. This confirms that the performance gains stem from the explicit logical and semantic structures modeled by pseudo-code.
\end{itemize}

\begin{tcolorbox}[
    colback=gray!5!white,
    colframe=gray!95!black,
    fonttitle=\bfseries,
    arc=2.8pt,
    boxrule=0.9pt,
    enhanced,
    boxsep=1.5pt,              
    left=1.5pt, right=1.5pt,     
    top=1.5pt, bottom=1.5pt,     
    before upper={\setstretch{0.9}}
]

\textbf{Finding 3}: PseudoBridge's superior performance in OOD scenarios demonstrates that its gains originate from the logical and semantic structure of pseudo-code rather than simple data scaling. By bridging NL intent and universal PL structures, pseudo-code provides invariant logical anchors that facilitate deep cross-modal alignment. This mechanism enables the model to learn generalizable logic instead of overfitting to surface-level syntactic patterns.
\end{tcolorbox}

\subsection{Qualitative Analysis}
\begin{figure}[htbp]
    \centering
    \includegraphics[width=\linewidth]{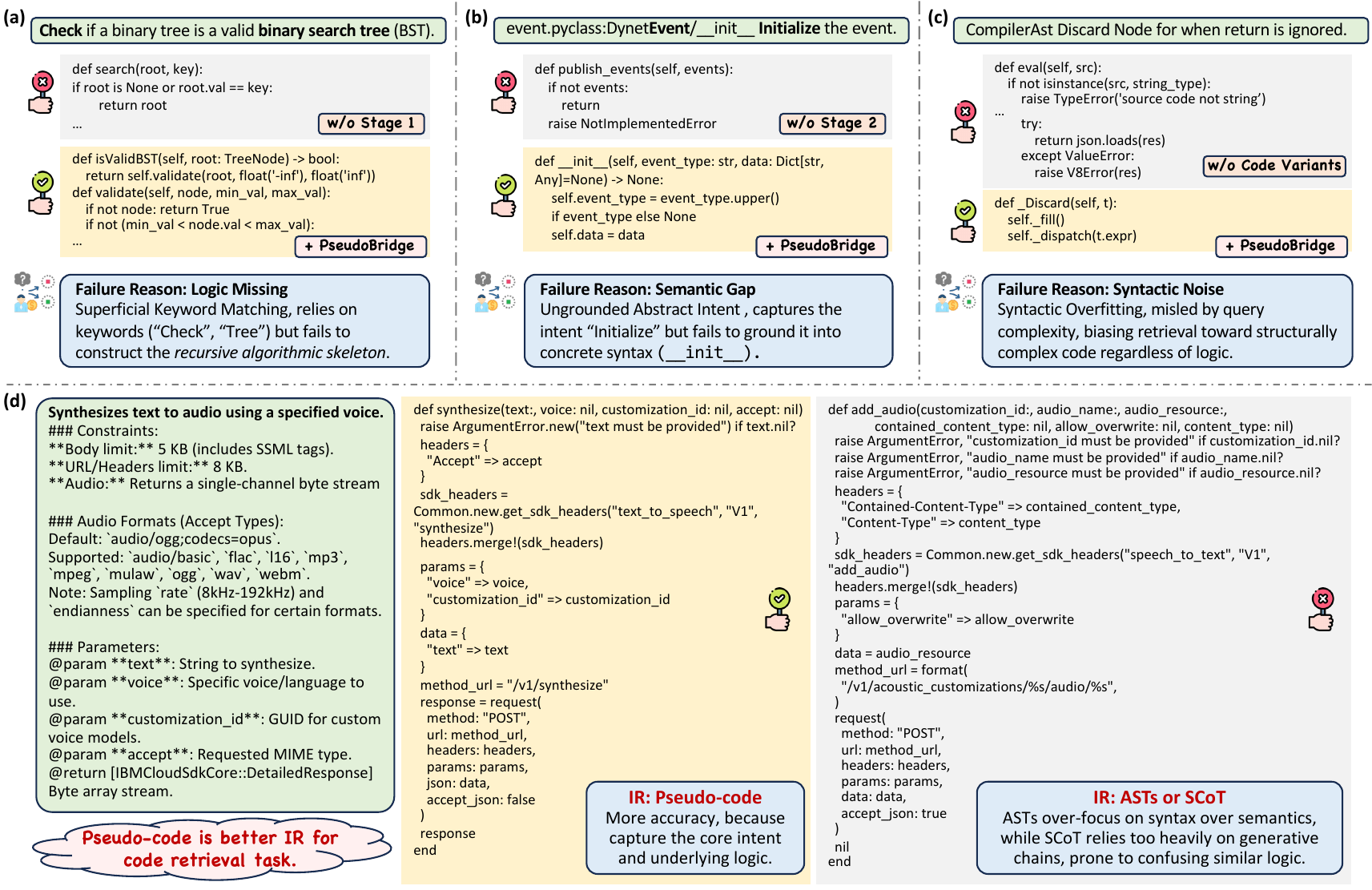}
    \caption{Qualitative analysis of retrieval cases showing the necessity of PseudoBridge components. (a) \textbf{\textit{w/o Stage 1}}: Fails to capture underlying code logic. (b) \textbf{\textit{w/o Stage 2}}: Fails to bridge query intent with implementation. (c) \textbf{\textit{w/o Code Variants}}: Relies on superficial heuristics, missing core semantics. (d) \textbf{IRs Comparison}: Pseudo-code outperforms AST (surface syntax) and SCoT (execution steps) by achieving deep alignment between NL intent and code.}
    \label{new_figs:case_study}
  \vspace{-0.15cm}
\end{figure}

To further explain the mechanisms behind the quantitative ablation results in Section 4.5, this section presents specific case studies. Figure~\ref{new_figs:case_study}
(a)-(c) qualitatively illustrate the practical effects of each component within the PseudoBridge framework. Figure~\ref{new_figs:case_study} (d) compares the performance of pseudo-code with other intermediate representations in code retrieval tasks.

%

\begin{itemize}
    \item \textbf{Case 1: Impact of <NL, Pseudo-code> Alignment.} Figure~\ref{new_figs:case_study} (a) illustrates a typical logic alignment error. In this failure case, the model captures keywords such as   ``Check'' and ``Tree''. However, it fails to comprehend that the core of verifying a BST lies in recursive constraint propagation. Due to the absence of pseudo-code generated by Stage 1 as structural support, the model fails to identify the correct recursive skeleton that includes the \texttt{helper(node, min, max)} function. This analysis indicates that the model is susceptible to the ambiguity of natural language without pseudo-code guidance. It degenerates into a keyword matching pattern. Therefore, the model struggles to capture algorithmic semantics and recursive logic.
    \item \textbf{Case 2: Impact of <Pseudo-code, Code> Alignment.} As shown in Figure~\ref{new_figs:case_study} (b), when the query explicitly requests the abstract concept of ``Initialize'', the model trained only with Stage 1 retrieves the functionally irrelevant \texttt{publish\_events} function. This result completely deviates from the initialization logic. This failure indicates a gap in the learning process. Although the model might comprehend the intention of `initialization'' through Stage 1, it lacks the mapping knowledge between pseudo-code and actual code established in Stage 2. Consequently, the model fails to grasp that ``Initialize'' corresponds to the specific \texttt{\_\_init\_\_} implementation in Python. As a result, the model reverts to a ``bag-of-words'' matching pattern and erroneously associates the keyword ``event'' with the query.
    \item \textbf{Case 3: Impact of Multi-Style Augmentation.} As shown in Figure~\ref{new_figs:case_study} (c), models without code variant augmentation tend to incorrectly retrieve structurally complex code when queries contain terms such as ``Compiler'' or ``AST''. This issue arises because complex queries frequently correspond to complex implementations in the training data. For instance, compiler-related code is typically lengthy. Consequently, the model acquires an erroneous heuristic that associates complex queries with complex syntax. Specifically, the model is attracted to syntactic structures like \texttt{try/except} blocks and \texttt{raise Error} statements. It simultaneously neglects the critical ``Discard'' semantics. By incorporating diverse code styles, the model learns to eliminate syntactic noise. It captures logical invariants directly. This approach improves the accuracy and robustness of code retrieval.
    \item \textbf{Case 4: Comparative Analysis of Intermediate Representations.} As shown in Figure~\ref{new_figs:case_study} (d), the user seeks to retrieve the \texttt{synthesize} method. However, both the ASTs and SCoT models incorrectly match the \texttt{add\_audio} function. This error arises from the high isomorphism in syntactic structures and control flows. Both functions exhibit typical API request construction patterns. In contrast, PseudoBridge leverages the de-syntaxization and de-proceduralization capabilities of pseudo-code. It aligns closely with the user's true intent and the underlying code logic. Consequently, PseudoBridge successfully returns the correct \texttt{synthesize} function. This demonstrates that pseudo-code surpasses surface syntax represented by ASTs and general execution steps represented by SCoT. It accurately captures the core logical invariant of text synthesis. Thus, the model achieves a deep alignment between the intent of natural language and the implementation of the code.
\end{itemize}

\section{Discussion}

\subsection{Analysis of Quality Threshold and Iterative Convergence}

\begin{figure}[htbp]
    \centering
    \includegraphics[width=\linewidth]{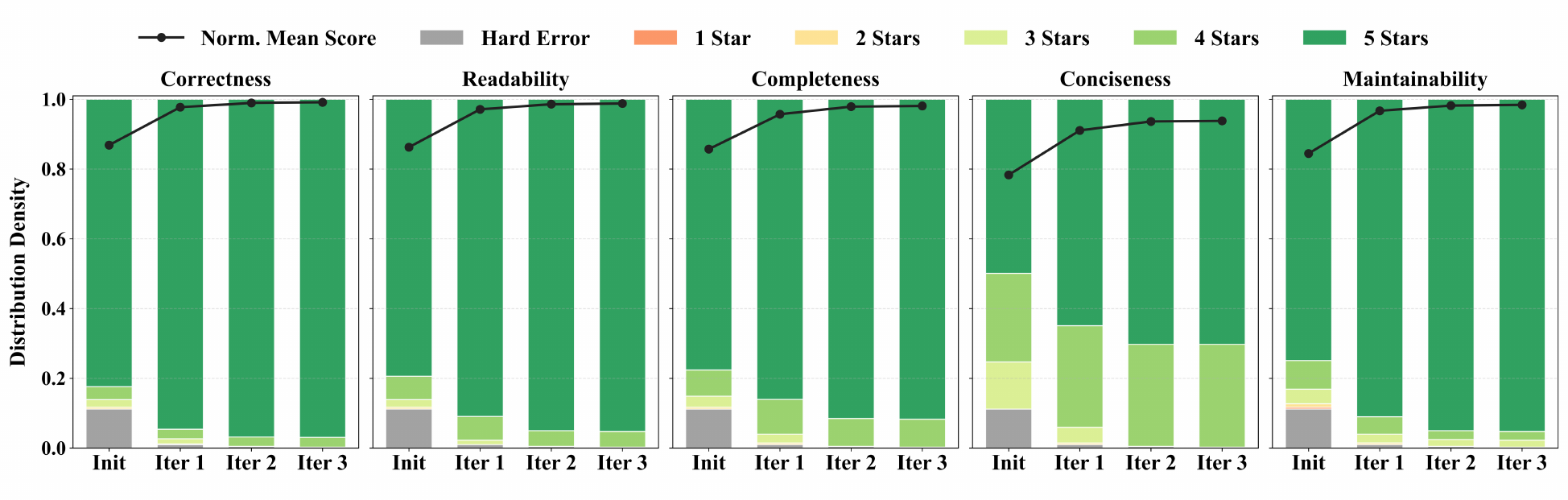}
    \caption{Pseudo-code quality score distributions across five dimensions over refinement iterations. Stacked bars represent score proportions (Hard Error to 5-Star), while the black line tracks the normalized mean score $[0, 1]$. Results show progressive quality improvement, with the most significant gains occurring after the first iteration.}
    \label{new_figs:pseudo_score_dis}
  \vspace{-0.15cm}
\end{figure}

To ensure the reliability of synthesized data, PseudoBridge establishes a quality evaluation system grounded in Likert Scale principles~\cite{likert1932technique}. These principles define a 4-star rating as the benchmark for high quality, representing substantial agreement or excellent performance within a five-star psychometric scale. Guided by this standard, our system evaluates quality across five dimensions, including correctness, readability, completeness, conciseness, and maintainability. Based on these criteria, this study sets the quality threshold at 4.0 to select high-quality samples that are suitable for downstream tasks.
Figure~\ref{new_figs:pseudo_score_dis} shows the distribution of quality scores during the iterative refinement process.

Statistics show that 11.46\% of samples do not reach the 4.0 threshold in the initial generation stage. Most defects come from logical inconsistency at 6.9\% and incomplete descriptions at 4.56\%. This distribution confirms that the automatic verification and iterative refinement mechanism is necessary. The score distribution moves toward the 4.0 to 5.0 range as the refinement process continues. The percentage of low-score samples drops significantly after the first iteration and the overall quality improves. Most unqualified samples reach the threshold within one or two iterations. The average score improves more slowly as the number of iterations increases. This convergence confirms that the 4.0 threshold effectively balances data quality and synthesis efficiency as a stable quality benchmark. This setting ensures that the large-scale corpus created by PseudoBridge maintains high fidelity for research and industrial applications.

\subsection{Cost Analysis}
Building on the quality convergence analysis above, we conduct a detailed cost analysis to evaluate the economic feasibility and scalability of the PseudoBridge framework.
Both stages incorporate an iterative \textbf{Evaluate and Refine} mechanism to ensure continuous quality improvement. 
For data generation and feedback-based refinement, we employ \textit{\textbf{GPT-4o}} (\textit{version:gpt-4o-2024-08-06}) to guarantee content accuracy. 
Conversely, for evaluation, we utilize the locally deployed \textit{\textbf{DeepSeek-R1-Distill-Llama-70B}} as the quality verifier. 
This strategy mitigates the potential self-preference bias associated with a single model family, thereby ensuring the objectivity and fairness of the assessment. 
Accounting for both the base generation investment and the dynamic overhead from the refinement phase, we define the total cost per sample as follows:

\begin{equation}\mathscr{C}_{total} = \underbrace{(\mathscr{I}_{gen} \cdot \mathscr{P}_{in} + \mathscr{O}_{gen} \cdot \mathscr{P}_{out})}_{\mathscr{C}_{base}} + \underbrace{E[R] \cdot (\mathscr{I}_{ref} \cdot \mathscr{P}_{in} + \mathscr{O}_{ref} \cdot \mathscr{P}_{out})}_{\mathscr{C}_{extra}}\end{equation}

where $\mathscr{C}_{total}$ represents the total cost incurred per sample. $\mathscr{C}_{base}$ denotes the initial generation cost, while $\mathscr{C}_{extra}$ accounts for the expected overhead from subsequent refinements. $\mathscr{P}_{in}$ and $\mathscr{P}_{out}$ are the unit prices for \textit{\textbf{GPT-4o}} Batch API ($\mathscr{P}_{in} = \$1.25/\text{1M}$, $\mathscr{P}_{out} = \$5.00/\text{1M}$). $\mathscr{I}_{gen}$ and $\mathscr{O}_{gen}$ indicate the token usage for the initial generation, whereas $\mathscr{I}_{ref}$ and $\mathscr{O}_{ref}$ represent the average token consumption during the refinement phase. Finally, $E[R]$ denotes the expected number of refinement iterations required to reach the quality threshold.

In the first stage, we focus on generating high-quality pseudo-code. We analyze the cost based on 1,000 randomly sampled data points. The cost is computed as follows:

\begin{itemize}
    \item \textbf{Generation:} Each sample consumes an average of 3,040 input tokens and 560 output tokens. The base cost is \textbf{$\mathscr{C}_{base} \approx \$0.0066$}.
    \item \textbf{Evaluation:} To minimize expenditures, we employ a locally deployed \textbf{\textit{DeepSeek-R1-Distill-Llama-70B}} model for logical equivalence checking and quality scoring. This local triage mechanism effectively eliminates external API calls during the verification phase.
    \item \textbf{Refinement:} The generation phase exhibits a comprehensive defect rate of 11.46\%. Upon triggering a refinement, the input context expands to approximately 3,304 tokens to incorporate diagnostic feedback. Given the decaying defect rate observed in subsequent iterations, the expected number of additional refinements per sample is $E[R] \approx 0.13$. Collectively, the expected refinement overhead is \textbf{$\mathscr{C}_{extra} \approx \$0.0010$}. 
\end{itemize}

For the second stage, we synthesize four distinct code variants for each validated pseudo-code sample. The cost analysis adheres to the same iterative logic:

\begin{itemize}
    \item \textbf{Generation:} To ensure stylistic diversity and structural complexity, the initial generation uses a prompt of 4,000 input tokens and yields 5,000 output tokens across all four variants, resulting in \textbf{$\mathscr{C}_{base} \approx \$0.0300$}.
    \item \textbf{Evaluation:} Each variant is independently verified by the local evaluator. Statistics indicate a first-pass success rate of 89.2\%. For the 10.8\% of variants failing verification, the system initiates a targeted refinement with an average of 3,200 input and 850 output tokens.
    \item \textbf{Refinement:} The expected number of refinement iterations per variant is 0.15. Since each sample consists of four variants, this leads to an average of $E[R] = 0.60$ refinement requests per sample. The resulting refinement overhead is \textbf{$\mathscr{C}_{extra} \approx \$0.0070$}.
\end{itemize}

By utilizing \textit{\textbf{GPT-4o}} and an evaluation mechanism driven by local LLMs, we minimize the total cost of data synthesis. 
The total cost of synthesizing a complete training sample, comprising high-quality pseudo-code and its four stylistic variants, is approximately \textbf{$\mathscr{C}_{total} \approx \$0.0446$}.
This competitive cost-efficiency makes the large-scale construction of high-quality program analysis datasets economically feasible. It further demonstrates the economic viability of the PseudoBridge framework for both industrial applications and academic research.

\begin{table}[htbp]
    \centering
    \caption{Retrieval time comparison (s) of different methods on Ruby and Python datasets. The best (lowest) times are highlighted in \textbf{bold}, demonstrating that adding PseudoBridge maintains or improves retrieval efficiency.}
    \label{new_tab:retrieval_efficiency}
    \resizebox{\textwidth}{!}{ 
        \begin{tabular}{c|c|cccccc}
            \toprule
            \multirow{2}{*}{\textbf{Dataset}} & \multirow{2}{*}{\textbf{Setting}} & \multicolumn{6}{c}{\textbf{Retrieval Time (s)}} \\
            \cmidrule(lr){3-8}
             & & \textbf{CodeBert} & \textbf{GraphCodeBert} & \textbf{DistilBert} & \textbf{UniXcoder} & \textbf{SentenceBert} & \textbf{CoCoSoDa} \\
            \midrule
            \multirow{2}{*}{\begin{tabular}{@{}c@{}}Ruby\\2,279\end{tabular}} 
             & Base & 10.20 & \textbf{10.30} & 4.93 & 9.65 & \textbf{9.32} & 9.15 \\
             & +PseudoBridge & \textbf{10.10} & 10.50 & \textbf{4.71} & \textbf{8.94} & 9.92 & \textbf{9.14} \\
            \midrule
            \multirow{2}{*}{\begin{tabular}{@{}c@{}}Python\\22,176\end{tabular}} 
             & Base & 209.53 & \textbf{226.90} & 396.66 & 249.16 & 247.80 & \textbf{191.76} \\
             & +PseudoBridge & \textbf{195.75} & 230.10 & \textbf{141.90} & \textbf{196.38} & \textbf{211.99} & 197.65 \\
            \bottomrule
        \end{tabular}
    }
\end{table}
\subsection{Practical Deployment and Efficiency}

To evaluate the universality and efficiency of PseudoBridge in practical applications, we conduct a comprehensive evaluation on three major categories of baselines: Encoder-only models (e.g., DistilBERT, CodeBERT), Unified and Generative models (e.g., UniXcoder), and Specific Retrieval frameworks (e.g., CoCoSoDa). 
The experimental results confirm that PseudoBridge has strong architectural compatibility.
It seamlessly enhances the semantic alignment capabilities of models ranging from general NLP models to advanced code retrieval models without changing the underlying bi-encoder structure. 
In terms of practical deployment, PseudoBridge adopts an ``offline enhancement, online zero-overhead'' strategy. 
The complex pseudo-code generation and intermediate modality alignment are completed entirely during the offline training phase. During online inference, the system only needs to store the optimized vector indices, without the need to cache text or perform generation tasks. Therefore, integrating this framework does not introduce any additional inference latency or storage burden. Especially for lightweight models like DistilBERT that prioritize speed, PseudoBridge significantly improves retrieval accuracy while perfectly maintaining millisecond-level response speeds (as shown in Table~\ref{new_tab:retrieval_efficiency}). This makes it suitable for direct deployment in industrial development environments, such as IDE plugins, which require high real-time performance.

\subsection{Potential for Other Downstream Tasks}

PseudoBridge demonstrates superior performance in code retrieval tasks. 
Furthermore, its core concept of utilizing pseudo-code as an intermediate modality for semantic and logical alignment holds significant potential for extension to other downstream software engineering tasks.

\subsubsection{Code Generation}
In the field of code generation, pseudo-code acts as a crucial intermediate link for chain-of-thought reasoning. 
Transforming complex natural language requirements into structured pseudo-code allows the model to explicitly plan algorithmic logic. 
This process effectively reduces the difficulty associated with directly generating target code. 
The progressive generation path moves from natural language to pseudo-code and then to specific implementation. 
This approach aligns closely with the alignment strategy constructed in this paper and helps improve the logical correctness and executability of the generated code.

\subsubsection{Code Summarization}
For code summarization, PseudoBridge enhances logical alignment, allowing models to bypass interference from syntactic details. The framework guides the model to focus on underlying algorithmic logic rather than surface-level styles. This capability enables the accurate retrieval of functional intent during the reverse transformation from PL to NL, ultimately producing concise and semantically precise descriptions.

\subsubsection{Code Translation}
Furthermore, pseudo-code operates as an intermediate representation independent of specific syntax, offering substantial benefits for code translation. The multi-style generation mechanism facilitates cross-language conversion by employing pseudo-code as a universal bridge. This process translates source code into an abstract logical form and subsequently reconstructs it in the target language. Consequently, this strategy masks syntactic differences and addresses the semantic gap effectively across diverse programming languages.

\subsection{Threats to Validity}

Although PseudoBridge effectively aligns natural language semantics with programming language logic to enhance code retrieval efficiency, the following potential challenges remain to be addressed in future work:

\subsubsection{Limited Scope of Evaluation}
Experimental evaluation builds upon established datasets and baselines, validating the approach across six representative programming languages and ten pre-trained models. 
These models cover both general-purpose and domain-specific fields. 
Although the results confirm the effectiveness of PseudoBridge, the performance variations across languages suggest that intrinsic properties, such as syntactic structures and programming paradigms, may influence the method's efficacy. 
Furthermore, our current evaluation does not yet fully cover emerging architectures designed for specific tasks, such as CodeBridge~\cite{liang2025codebridge} for zero-shot code understanding. 
Therefore, future work aims to expand the range of programming languages and systematically explore the transferability of our framework to a broader set of architectures. 
This will help us comprehensively verify its universality and robustness within complex software ecosystems.

\subsubsection{Dependence on LLMs Reliability}
Experimental results demonstrate that PseudoBridge exhibits strong robustness across various generation models, ranging from \textit{StarCoder2} to \textit{GPT-4o}. However, the upper bound of its retrieval performance remains fundamentally constrained by the instruction-following and reasoning capabilities of the underlying LLMs. Due to the probabilistic nature of LLMs, errors inevitably accumulate during generation. Although we employ an \textbf{Evaluate and Refine} mechanism to filter out low-quality outputs, this process depends on the discriminative capability of the model itself and thus does not fully eliminate error propagation. To address these limitations, our future work focuses on two main directions. First, we explore fine-tuning small, specialized models for pseudo-code generation to maintain high synthesis quality while reducing deployment costs. Second, we integrate advanced reinforcement learning methods, such as GRPO~\cite{guo2025deepseekr1} and DAPO~\cite{yu2025dapo}, into the generation pipeline. By designing effective reward functions, we aim to enhance the reliability and accuracy of the intermediate representations.

\subsubsection{Distributional Bias in Generation}

Relying on LLMs for data generation, PseudoBridge inevitably inherits biases favoring mainstream languages like Python and Java. We address this via strict prompt engineering, enforcing standardized structures for pseudo-code and multi-dimensional style constraints for code variants. These measures decouple syntax dependencies and promote the exploration of long-tail distributions, ensuring the neutrality of intermediate representations. While explicit bias decreases, implicit semantic tendencies persist. Future work therefore prioritizes two directions: integrating diverse LLM families to balance individual model biases, and establishing fine-grained benchmarks to assess the semantic fidelity of low-resource languages. By quantifying these discrepancies, we enhance the framework's fairness and cross-language adaptability.

\section{Conclusion}
In this paper, we propose PseudoBridge, a novel code retrieval framework leveraging pseudo-code as an intermediate semi-structured modality, enabling more precise alignment between NL semantics and PL logic. 
In addition, we design a logic-invariant code style augmentation strategy, using LLMs to generate code implementations that are stylistically diverse yet logically equivalent and aligned with the same pseudo-code. 
We perform extensive evaluations across ten baselines and six mainstream programming languages, confirming substantial improvements in retrieval accuracy with PseudoBridge enhancement.
Results show that PseudoBridge consistently outperforms existing methods, particularly in challenging zero-shot domain transfer scenarios such as Solidity and XLCoST datasets.
Future research will conduct tests on more programming languages and compare the impact of different advanced LLMs on the framework.

\section{Data Availability}
Our source code and synthesized data are available at \url{https://github.com/yixuanli1230/PseudoBridge}.



\bibliographystyle{ACM-Reference-Format}
\bibliography{ref}

\appendix

\section{Prompt Template}
\label{appendix:a}

\begin{figure*}[ht]
    \centering
    \includegraphics[width=0.85\linewidth]{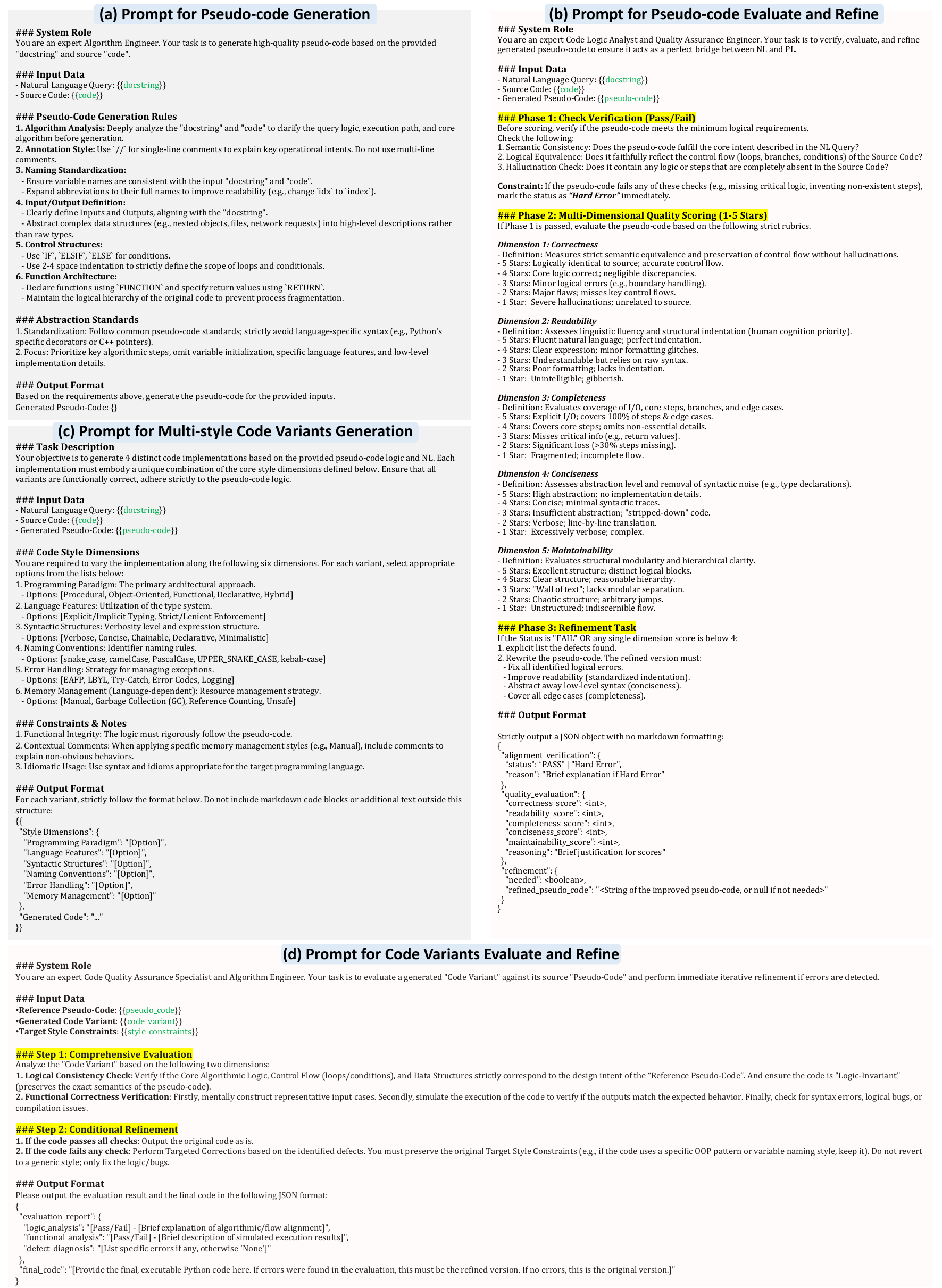}
    \caption{Overview of the prompt templates used in the PseudoBridge data synthesis pipeline. The pipeline consists of four sequential stages: (a) generating pseudo-code from NL and PL; (b) evaluating and iteratively refining the generated pseudo-code; (c) synthesizing multi-style code variants based on the refined pseudo-code; and (d) verifying and refining the functional correctness of the generated code variants.}
    \label{new_fig:prompt_a}
    \vspace{-0.15cm}
\end{figure*}

Figure~\ref{new_fig:prompt_a} details the prompt template process for data synthesis within the PseudoBridge framework. This process systematically constructs and verifies high-quality pseudo-code and multi-style code variants from natural language and source code through four ordered stages:

\begin{itemize}
    \item \textbf{(a) Pseudo-code Generation:} This phase corresponds to \textbf{Figure~\ref{fig:PseudoBridge_overview} Step 1}. It aims to construct a semantic bridge between NL and PL. The prompt template configures the system to function as an algorithm engineer that accepts an NL query and source code as inputs. The system enforces strict generation rules involving algorithmic logic analysis, namespace standardization, and control flow specifications. It also applies high-level abstraction criteria. Consequently, the model filters out language-specific implementation details. It produces a standardized pseudo-code intermediate representation that preserves core algorithmic logic while ensuring high readability.
    \item \textbf{(b) Pseudo-code Evaluation and Refinement:} As illustrated in \textbf{Figure~\ref{fig:PseudoBridge_overview} Step 2}, this phase introduces an ``Evaluate and Refine'' closed-loop mechanism to ensure the reliability of pseudo-code quality. The system first executes a binary verification (Pass/Fail) based on logical equivalence and hallucination detection. Subsequently, for samples that pass verification, the system conducts a quantitative scoring (1-5 stars) across five dimensions: correctness, readability, completeness, conciseness, and maintainability. If a sample fails to meet the threshold in any dimension, the system triggers a targeted rewriting task based on the diagnosed defects, thereby achieving iterative optimization of pseudo-code quality.
    \item \textbf{(c) Multi-style Code Variants Generation:} This phase aligns with \textbf{Figure~\ref{fig:PseudoBridge_overview} Step 3}. It focuses on expanding structural diversity while preserving logical invariance. The prompting framework establishes a style constraint space with six core dimensions, including programming paradigms, error handling strategies, and memory management methods. Based on high-quality pseudo-code, the model generates multiple code implementations. These variants are functionally equivalent but syntactically distinct. This process constructs a robust training dataset characterized by invariant logic and diverse styles.
    \item \textbf{(d) Code Variants Evaluate and Refine:} The final stage corresponds to \textbf{Figure~\ref{fig:PseudoBridge_overview} Step 4} and is responsible for ensuring the functional correctness of the generated multi-style code. The system acts as a quality assurance specialist and performs a dual verification on the code variants. First, it executes a logical consistency check to ensure that the algorithmic intent remains unchanged. Second, it constructs test cases through mental simulation to verify functional correctness. For code that fails verification, the system performs targeted corrections while strictly maintaining the original style constraints. This process ensures that every sample in the final dataset possesses both stylistic diversity and functional accuracy.
\end{itemize}

\section{Case Study}

\subsection{High-quality Pseudo-code Generation}
\label{sec5.1.1}
\begin{figure*}[ht]
    \centering
    \includegraphics[width=0.9\linewidth]{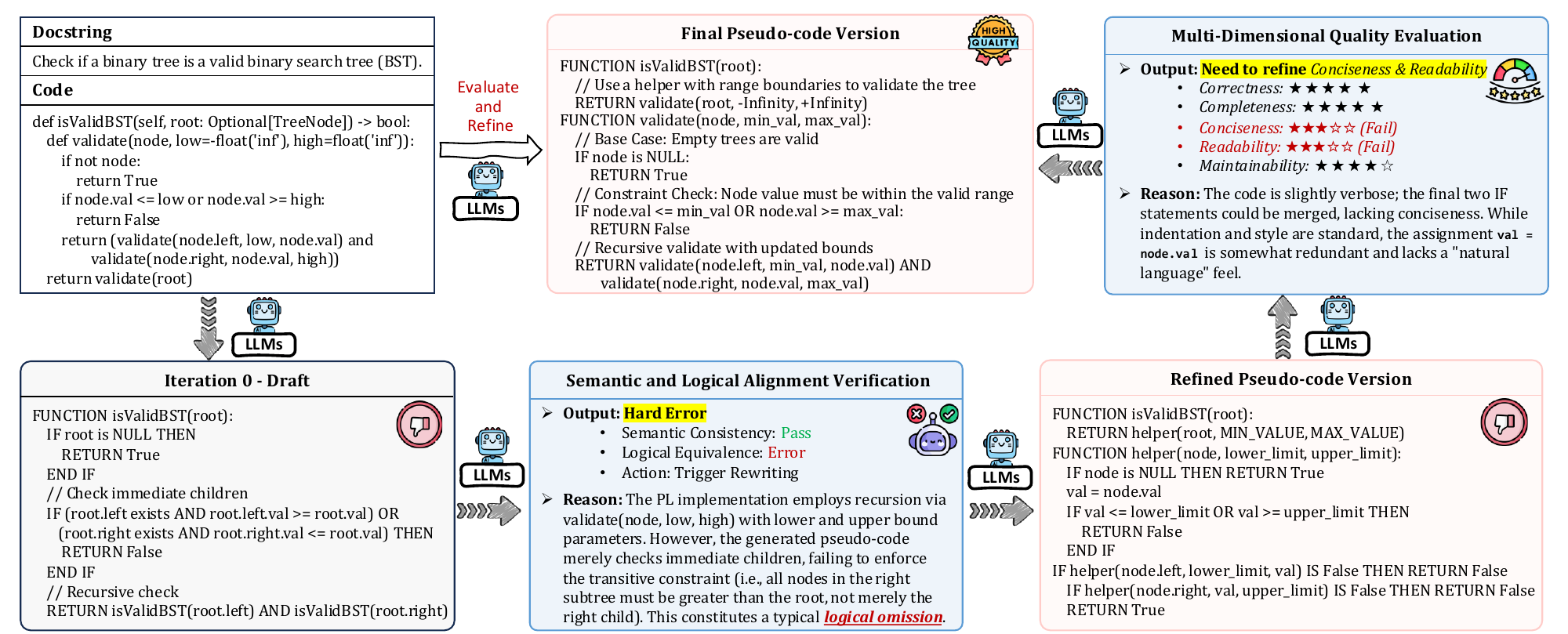}
    \caption{Examples of high-quality pseudo-code generation.}
    \label{new_fig:pseudocode_quality}
    \vspace{-0.15cm}
\end{figure*}

Figure~\ref{new_fig:pseudocode_quality} illustrates the PseudoBridge process using a Binary Search Tree (BST) validation example. The original code utilizes a recursive function with boundary constraints. In the initial generation, the LLM failed to propagate recursive boundary conditions, performing only local checks on immediate child nodes. The system identified this critical flaw during verification, classifying it as a \textbf{``Hard error''} and triggering a rewrite. The subsequent iteration corrected the logic but introduced redundant checks and verbosity. Consequently, it was penalized for poor readability and rejected for falling below the acceptance threshold. Finally, guided by evaluation feedback, the system generated a refined version featuring clear logical operators and intuitive boundary descriptions, successfully meeting all quality criteria.

\subsection{Multi-style Code Variants Generation}
\label{sec5.1.2}

\begin{figure*}[ht]
    \centering
    \includegraphics[width=0.9\linewidth]{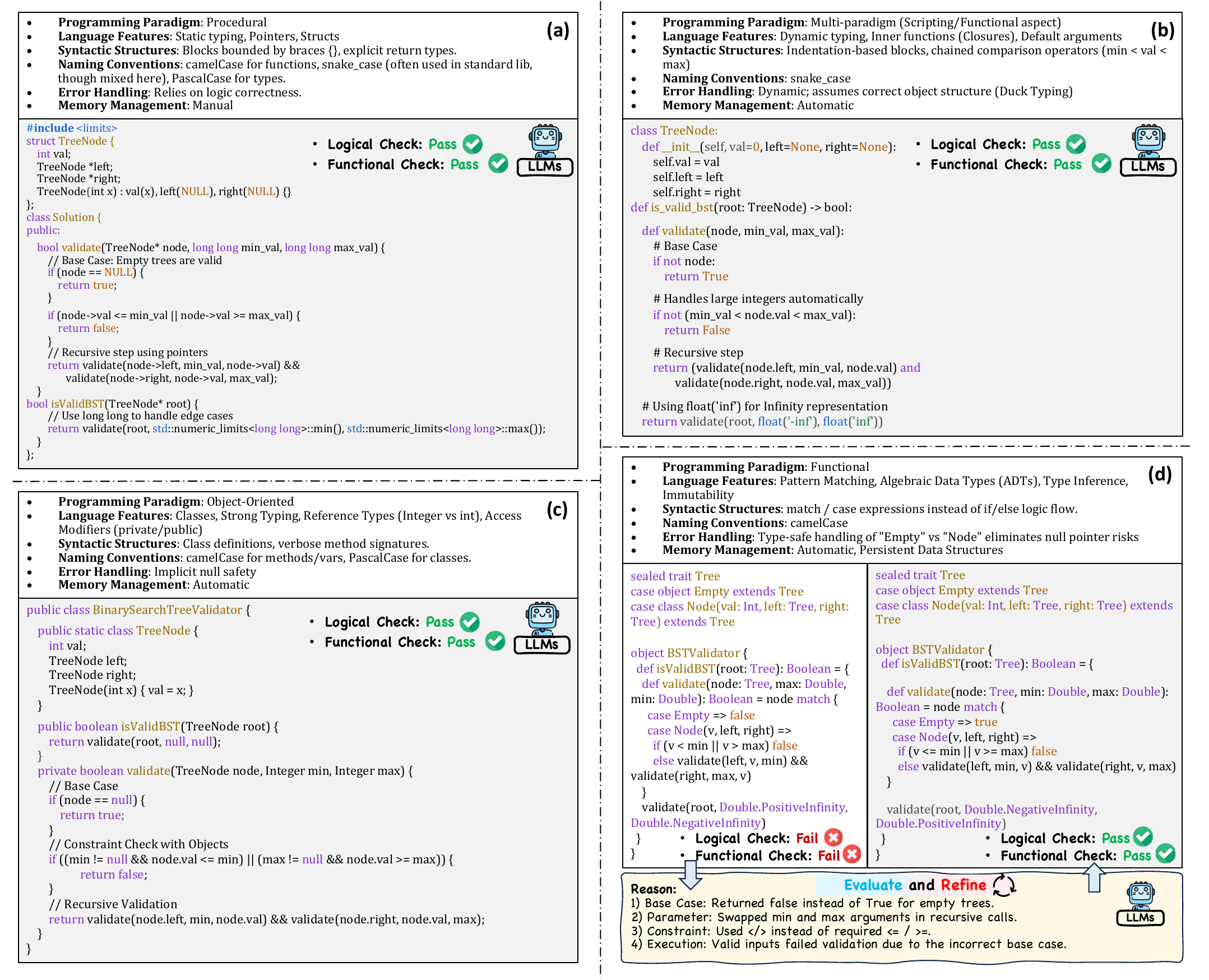}
    \caption{Examples of multi-style code generation based on high-quality pseudo-code. (a) C++ (Procedural): Focuses on manual control and pointer-based efficiency. (b) Python (Concise/Functional): Prioritizes readability and dynamic logic expression. (c) Java (Object-Oriented): Emphasizes encapsulation, type safety, and robust exception handling. (d) Scala (Functional): Utilizes immutability and pattern matching, replacing control flow statements with recursive definitions.}
    \label{new_fig:mutlicode_gen}
    \vspace{-0.15cm}
\end{figure*}

To demonstrate the effectiveness of our approach, we generated four diverse code implementations for BST validation based on identical pseudo-code, as shown in Figure~\ref{new_fig:mutlicode_gen}: (a) a C++ version emphasizing low-level efficiency; (b) a Python version prioritizing logical conciseness; (c) a Java version reflecting strict object-oriented encapsulation; and (d) a Scala version utilizing functional pattern matching.
We document a typical refinement process using the Scala implementation (Figure~\ref{new_fig:mutlicode_gen} (d)). The initial output fails validation due to logical errors, including incorrect base case handling, swapped recursive parameters, and imprecise comparison operators. Guided by this feedback, the system initiates a refinement cycle. The model successfully rectifies these issues by fixing the return value, adjusting the parameter order, and enforcing strict comparisons. This process ultimately yields a correct implementation that retains its distinct functional style.

\section{Proprietary LLM Parameters for Build Data}
\label{appendix_llm_parameters}
The detailed parameters for the data synthesis pipeline using \textit{\textbf{GPT-4o}} and the automated quality assurance using \textit{\textbf{DeepSeek-R1}} are listed in Table~\ref{tab:llm_parameters}. We differentiate the temperature settings for pseudo-code and multi-style code generation to balance stability and diversity.

\begin{table}[htbp]
\centering
\caption{Hyper-parameter configurations for the three phases of PseudoBridge.}
\label{tab:llm_parameters}
\begin{tabularx}{\textwidth}{l|c|c|c}
\toprule[1.0pt]
\textbf{Parameters} & \textbf{Pseudo-code Generation} & \textbf{Multi-style Code Generation} & \textbf{Automated Evaluation} \\ 
\midrule
Model Version       & gpt-4o-2024-08-06 & gpt-4o-2024-08-06 & deepseek-r1-distill-llama-70b \\
Temperature         & 0.6               & 0.8               & 0.2                           \\
Max\_tokens         & 2,048              & 4,096              & 8,192                          \\
n                   & 1                 & 1                 & 1                             \\
Top\_p              & 1.0               & 1.0               & 1.0                           \\
Frequency\_penalty  & 0                 & 0                 & 0                             \\
Presence\_penalty   & 0                 & 0                 & 0                             \\ 
\bottomrule[1.0pt]
\end{tabularx}
\end{table}

\section{Hyperparameters for Model Tuning}
\label{appendix_finetuning_parameters}
The fine-tuning of retrieval models is conducted on two NVIDIA A100-SXM4 GPUs. The detailed hyperparameters used for the two-stage fine-tuning process are presented in Table~\ref{tab:finetuning_parameters}. Note that these settings are applied consistently across different model architectures (e.g., CodeBERT, UniXcoder).

\begin{table}[h]
\centering
\caption{Hyperparameter settings for retrieval model fine-tuning.}
\label{tab:finetuning_parameters}
\begin{tabular}{l|c||l|c||l|c}
\toprule[1.0pt]
\textbf{Hyperparameter} & \textbf{Value} & \textbf{Hyperparameter} & \textbf{Value} & \textbf{Hyperparameter} & \textbf{Value} \\ \midrule
Optimizer       & AdamW  & Max. Context Length & 1,024 & Training Precision & FP32 \\
Learning Rate   & 5e-5   & Num. Epochs         & 3     & Weight Decay       & 0.01 \\
Warmup Ratio    & 0.1    & Batch Size          & 48    & Pooling Mode       & CLS Token \\
LR Scheduler    & Linear & Random Seed         & 42    &                    & \\ \bottomrule[1.0pt]
\end{tabular}
\end{table}






\end{document}